\documentclass[%
reprint,
superscriptaddress,
showpacs,
nofootinbib,
amsmath,amssymb,
prc,
floatfix, ]%
{revtex4-1}

\usepackage{color}

\usepackage{graphicx}% Include figure files
\usepackage{dcolumn}% Align table columns on decimal point
\usepackage{bm}% bold math
\usepackage[dvipdfmx,bookmarks=true,colorlinks,%
            citecolor=blue,linkcolor=blue,anchorcolor=blue,filecolor=blue,urlcolor=blue,%
           ]{hyperref}
\allowdisplaybreaks

\newcommand{\bra}[1]{\langle {#1} |}
\newcommand{\ket}[1]{| {#1} \rangle}

\begin{document}

%\begin{CJK*}{GBK}{}

\title{
Adiabatic self-consistent collective path in nuclear fusion reactions}

\author{Kai Wen}%
 \email{wenkai@nucl.ph.tsukuba.ac.jp}
 \affiliation{Center for Computational Sciences,
              University of Tsukuba, Tsukuba 305-8577, Japan}

\author{Takashi Nakatsukasa}%
 \affiliation{Center for Computational Sciences,
              University of Tsukuba, Tsukuba 305-8577, Japan}
 \affiliation{Faculty of Pure and Applied Sciences,
              University of Tsukuba, Tsukuba 305-8571, Japan}
 \affiliation{iTHES Research Group, RIKEN, Wako 351-0198, Japan}

\date{\today}

\begin{abstract}
Collective reaction paths for fusion reactions,
$^{16}$O+$\alpha$ $\rightarrow$ $^{20}$Ne
and $^{16}$O+$^{16}$O $\rightarrow$ $^{32}$S,
are microscopically determined,
on the basis of the adiabatic self-consistent collective coordinate (ASCC)
method.
The collective path is maximally decoupled from other intrinsic degrees of
freedom.
The reaction paths turn out to deviate from those obtained with standard
mean-field calculations with constraints on
quadrupole and octupole moments.
The potentials and inertial masses defined in the ASCC method
are calculated along the reaction paths, which leads to
the collective Hamiltonian used for calculation of
the sub-barrier fusion cross sections.
The inertial mass inside the Coulomb barrier may
have a significant influence on the fusion cross section at
the deep sub-barrier energy.
\end{abstract}


\maketitle

%\end{CJK*}

\section{Introduction}
The microscopic description of the large amplitude nuclear collective motion is one
of the major and long-standing problems in nuclear physics.
For the collective motion in a complex multinucleon system,
%As a problem of many-body dynamics, theoretically
it is useful to describe its dynamics in terms of
a small number of collective coordinates $q$.
In most cases, one ``intuitively'' adopts
deformation parameters $q$, such as the intrinsic quadrupole moment.
In the energy density functional (EDF) approaches \cite{SR16},
a collective subspace (path) $|\psi(q)\rangle$
is constructed by performing
the constrained minimization calculation with one-body
constraining operators associated with those deformation parameters.
Even simpler methods could be adopted,
by assuming a single-particle potential, such as the Nilsson potential,
as a function of the deformation parameters \cite{Bra72}.
%When we introduce the collective coordinate $q$ ,
%we've been submitted to the conception of classical equation of motion,
It is certainly desirable
to microscopically
extract a few collective variables,
without relying on our intuitive choice,
which are maximally decoupled from all the other
intrinsic degrees of freedom.
%whose equation of motion
%obeys the classical Hamilton's equations of motion,
%these canonical variables are regarded as the optimal collective
%variables

The most well-known theory for this purpose,
to determine such an optimal collective subspace,
is the adiabatic time-dependent Hartree-Fock  (ATDHF) theory
\cite{BGV76,Vil77,BV78,GR78}.
The ATDHF theory is derived by using the expansion with respect to
collective momenta $p$ up to the first order.
In practical calculations, the ATDHF is formulated into the form of
differential equation with initial states.
Starting from different initial conditions, the ATDHF
equation produces different collective paths.
Many trajectories must be produced to find the ``best'' one
\cite{GGR83,RG87}.
This is known as a  ``non-uniqueness'' problem.
A possible way to overcome this problem is to
take into account the second-order terms in momentum $p$
\cite{MP82,DKW00}.

%In the adiabatic limit, assuming the collective motion is slow,
The adiabatic self-consistent collective coordinate (ASCC) method
provides us an alternative approach, free from the non-uniqueness problem,
to determining the optimal collective path or a collective sub-manifold
embedded in the large-dimensional  phase space of Slater determinants
\cite{MNM00,NMMY16}.
The ASCC method has been applied to nuclear structure problems
\cite{HNMM08,HNMM09,HSNMM10,HSYNMM11,
HLNNV12,SHYNMM12,MMNYHS16,Nak12}.
In Ref.~\cite{WN16}, we have proposed a numerical method to solve the ASCC
equations for nuclear reaction,
combining the imaginary-time
evolution \cite{DFKW80} and the finite amplitude method \cite{NIY07,AN11,AN13}.
The test calculation has been done for the simplest system of the reaction path
of $\alpha$+$\alpha$ $\rightarrow$ $^{8}$Be. %, the corresponding inertial mass
%parameter along the collective coordinate can also be calculated with microscopic
%basis.
The present paper is on the continuation of this work.
The numerical methods proposed in Ref.~\cite{WN16} are
applied to nuclear fusion reactions of
$^{16}$O+$\alpha$ $\rightarrow$ $^{20}$Ne
and $^{16}$O+$^{16}$O $\rightarrow$ $^{32}$S.
%$^{16}$O $\leftrightarrow$ $^{12}$C+$\alpha$,
%$^{20}$Ne $\leftrightarrow$$^{16}$O+$\alpha$,
%$^{32}$S $\leftrightarrow$ $^{16}$O+$^{16}$O.
They demonstrate unique features of the collective dynamics,
showing that the optimal collective path can be different
from the constrained Hartree-Fock (CHF) states with constraints on
the mass quadrupole and octupole moments.

The inertial mass is another important issue in nuclear collective motion.
The ASCC method is capable of providing the masses
for the collective motion in the decoupled
subspace including effects of time-odd mean fields.
They are different from other known inertial masses, such as those of
the gaussian overlap approximation for the generator coordinate method
and the cranking formula \cite{RS80}.
We will show significant difference between the ASCC and the cranking
formulae, especially inside the Coulomb barrier.

The paper is organized as follows.
In Sec.~\ref{sec:theo}, we give the formulation of the basic ASCC
equations, %in the case of one-dimensional collective motion,
to determine the collective path and
%the coordinate transformation procedure
to calculate the mass parameter.
In Sec.~\ref{sec:path}, we apply the method to extract the collective
paths for the reaction systems of
$^{16}$O+$\alpha$ $\rightarrow$ $^{20}$Ne
and $^{16}$O+$^{16}$O $\leftrightarrow$ $^{32}$S.
The inertial mass with respect to the relative
distance between two nuclei are calculated.
The sub-barrier fusion cross section
%for $^{16}$O+$^{16}$O $\leftrightarrow$ $^{32}$S
is estimated from the results.
Summary and concluding remarks are given in In Sec.~\ref{sec:sum}

%%%%%%%%%%%%%%%%%%%%%%%%%%%%%%%%%%%%%%%%%%%%%%%%%%%%%%%%%%%%%%%%%%%%
\section{\label{sec:theo}Theoretical framework}

%%%%%%%%%%%%%%%%%%%%%%%%%%%%%%%%%%%%%%%%%%%%%%%%%%%%%%%%%%%%%%%%%%%%
In this section, we recapitulate the basic formulation of the
ASCC method without the pairing correlation.
Then, we briefly describe a procedure to construct the
one-dimensional (1D) collective
path and to calculate the inertial mass.
The details can be found in Ref.~\cite{WN16}.

%%%%%%%%%%%%%%%%%%%%%%%%%%%%%%%%%%%%%%%%%%%%%%%%%%%%%%%%%%%%%%%%%%%%
\subsection{\label{subsec:theoA}Basic equations of the ASCC method}

In the present study, we assume that the reaction is described
by the 1D collective coordinate $q(t)$
and its conjugate momentum $p(t)$.
Parameterizing the time-dependent mean-field states (Slater determinants)
as $\ket{\psi(p,q)}$,
the total energy of the system in this parametrization reads
\begin{eqnarray}
H(p, q)= \langle \psi(p, q)|\hat{H} |\psi(p, q)\rangle, \label{pdef}
\end{eqnarray}
which defines a classical collective Hamiltonian.
In the ASCC method,
the optimal collective path $\ket{\psi(p,q)}$ is obtained so as to
be maximally decoupled from the intrinsic degrees of freedom.
Therefore,
the evolution of $q(t)$ and $p(t)$ approximately
obey the canonical equation of motion
with the Hamiltonian $H(p, q)$.

The state $\ket{\psi(p,q)}$ is written in powers of $p$ about $p = 0$ as
\begin{eqnarray}
|\psi(p,q)\rangle = e^{i p \hat{Q}(q)}|\psi(0,q)\rangle
	= e^{i p \hat{Q}(q)}\ket{\psi(q)},
\label{eq1-2}
\end{eqnarray}
where $\hat{Q}(q)$ is defined as
$\hat{Q}(q)|\psi(q)\rangle = -i \partial_{p}|\psi(q)\rangle$.
The conjugate operator $\hat{P}(q)$ is introduced as an infinitesimal
generator for translating the system with respect to $q$,
$\hat{P}(q)|\psi(q)\rangle = i \partial_{q}|\psi(q)\rangle$.
$\hat{P}(q)$ and $\hat{Q}(q)$
can be expressed in the form of one-body operator as
\begin{eqnarray}
\hat{P}(q)&=&i \sum_{n\in p,j\in h}P_{nj}(q)a^{\dag}_{n}(q)a_{j}(q)
+ \mathrm{h.c.}, \label{P2}\\
\hat{Q}(q)&=&\sum_{n\in p,j\in h}Q_{nj}(q)a^{\dag}_{n}(q)a_{j}(q)
+ \mathrm{h.c.}.
 \label{Q(q)}
\end{eqnarray}
They are locally defined at each $q$ and change their
structure along the collective path.
The particle ($n\in p$) and hole ($j\in h$) states are also defined with
respect to $\ket{\psi(q)}$.
The weak canonicity condition
\begin{eqnarray}
 \langle\psi(q)|[i\hat{P}(q),\hat{Q}(q)]|\psi(q)\rangle=1 \label{weak}
\end{eqnarray}
is imposed to make $q$ and $p$ a pair of canonical variables.

The self-consistent collective coordinate (SCC) method is based on
the invariance principle of the time-dependent mean-field theory \cite{MMSK80}.
%In the adiabatic limit, the collective momentum $p$ is assumed to be small, % the collective motion
The adiabatic approximation in ASCC refers to the
assumption that the collective momentum $p$ is small,
so that we can expand equations in terms of $p$
up to the order of $p^{2}$.
The invariance principle of SCC
leads to the following set of ASCC equations
\cite{MNM00,NMMY16},
\begin{eqnarray}
&&\delta \langle{\psi(q)} |\hat{H}_{\rm mv}(q)|{\psi(q)}\rangle = 0
 \label{chf} \\
&&\delta\langle \psi(q)|[\hat{H}_{\rm mv}(q),\frac{1}{i}\hat{P}(q) ]
        - \frac{\partial^{2} V}{\partial q^{2}} \hat{Q}(q)
        |\psi(q)\rangle = 0,
 \label{ASCC1} \\
&&\delta\langle \psi(q)|[\hat{H}_{\rm mv}(q),i\hat{Q}(q)]
         - \frac{1}{M(q)}\hat{P}(q)   |\psi(q)\rangle = 0,
\label{ASCC2}
\end{eqnarray}
where $\hat{H}_{\rm mv}(q)\equiv \hat{H}-(\partial V/\partial q)\hat{Q}(q)$
is the ``moving'' Hamiltonian.
Here, the curvature term, associated with $d\hat{Q}/dq$,
is neglected for simplicity \cite{NMMY16}.
The collective potential $V(q)$ is defined as
\begin{eqnarray}
V(q)= \langle \psi(q)|\hat{H} |\psi(q)\rangle,
\label{vdef}
\end{eqnarray}
and $M(q)$ is the inertial mass of the collective motion.
Equation (\ref{chf}) is called
``moving mean-field equation'' (``moving Hartree-Fock (HF) equation''),
and Eqs. (\ref{ASCC1}) and (\ref{ASCC2})
are ``moving random-phase approximation (RPA)''.
This set of equations determines the reaction path
 $\ket{\psi(q)}$ as well as the local generators,
$\hat{P}(q)$ and $\hat{Q}(q)$, self-consistently.

To fix the scale of $\hat{P}(q)$ and $\hat{Q}(q)$,
for the present study,
we set the mass $M(q)$ in Eq. (\ref{ASCC2}) to be a constant value,
$M(q)=M_q=1\ \hbar^2$MeV$^{-1}$fm$^{-2}$.
%The operators $\hat{Q}(q)$ and $\hat{P}(q)$
%are determined by the moving RPA Eqs. (\ref{ASCC1}) and (\ref{ASCC2}),
%and the wave function $\psi(q)$ %in Eqs. (\ref{ASCC1}) and (\ref{ASCC2})
%is constructed by Eq. (\ref{chf}) with $\hat{H}_{\rm mv}(q)$.
%At $\psi(q)$, the local operator for generating the moving RPA collective excitation
%state $\Omega^{\dag}(q)|0\rangle$ can be constructed as
%\begin{equation}
%\Omega^\dagger(q)=\sqrt{\frac{\omega(q)}{2}} \hat{Q}(q)
%-\frac{i}{\sqrt{2\omega(q)}} \hat{P}(q),
%\label{omega}
%\end{equation}
%where $\omega$ is the eigenfrequency of the moving RPA
%equations (\ref{ASCC1}) and (\ref{ASCC2}).
%The square of the moving RPA eigenvalue $\omega^{2}$
%Eqs. (\ref{ASCC1}) and (\ref{ASCC2})
This determines the scale and the dimension of the coordinate $q$.
The second order derivative of the potential energy with respect to $q$
corresponds to the squared frequency of the moving RPA.
\begin{eqnarray}
\omega^{2}(q)=\frac{1}{M_q}\frac{\partial^{2} V}{\partial q^{2}} .
%            =\frac{\partial^{2} V}{\partial q^{2}}.
	\label{ab4}
\end{eqnarray}

%where $\hat{P}(q)$ and $\hat{Q}(q)$ are the solutions of Eqs. (\ref{ASCC1}) and (\ref{ASCC2})
%based on the state $\psi(q)$,
%the local forward and backward amplitude $X_{\rm ni}(q)$ and
%$Y_{\rm ni}(q)$ of the RPA collective mode are the linear combination
%of $P_{\rm ni}(q)$ and $Q_{\rm ni}(q)$
%\begin{eqnarray}
%   X_{nj}(q) &=&\sqrt{\frac{\omega}{2}}Q_{nj}(q)
%     +\frac{1}{\sqrt{2\omega}}P_{nj}(q),\nonumber \\
%   Y_{nj}(q) &=&\sqrt{\frac{\omega}{2}}Q_{nj}(q)
%     -\frac{1}{\sqrt{2\omega}}P_{nj}(q). \\
%\end{eqnarray}
%where the indexes $i, j$ and $n,m$ refer to the hole and particle states
%respectively.

%which is equivalent to the RPA normalization condition
%\begin{eqnarray}
 %\sum_{n,j}(X_{nj}^{2}(q)-Y_{nj}^{2}(q))=1.
%\end{eqnarray}

To solve the moving RPA equations (\ref{ASCC1}) and (\ref{ASCC2}),
we make use of the the finite amplitude method (FAM)
~\cite{Nakatsukasa2007_PRC76-024318,
Avogadro2011_PRC84-214314},
especially the matrix FAM prescription~\cite{Avogadro2013_PRC87-014331}.
In the FAM,
only the calculations of the single-particle Hamiltonian
constructed with independent bra and ket states are required
~\cite{Nakatsukasa2007_PRC76-024318},
providing us a high numerical efficiency to solve
Eqs. (\ref{ASCC1}) and (\ref{ASCC2}).
The moving mean-field equation (\ref{chf}) is solved by using the
imaginary-time method.
In practical calculation, we adopt
the coordinate-space representation for the mean-field states
and the mixed representation for the
RPA matrix \cite{WN16}.

%%%%%%%%%%%%%%%%%%%%%%%%%%%%%%%%%%%%%%%%%%%%%%%%%%%%%%%%%%%%%%%%%%%%
\subsection{\label{sec:theob}Construction of collective reaction path}

We may start the construction of the collective path,
in principle, from any state $\ket{\psi(q)}$ that satisfies
Eqs. (\ref{chf}), (\ref{ASCC1}), and (\ref{ASCC2}).
There are a kind of ``trivial'' states;
the ground state of the whole system (after fusion) and
the state with well separated projectile and target (before fusion).
We start the construction procedure from one of these trivial
initial states.
At the initial state $\ket{\psi(q=0)}$ on the collective path,
the solutions of the moving RPA equations (\ref{ASCC1}) and (\ref{ASCC2})
provide many kinds of modes,
among which we need to select the one associated with the reaction path.
Here, we choose the lowest mode of excitation except for the
Nambu-Goldstone (NG) modes associated with
the translation and rotation of the total system.

To identify character of the modes,
we calculate the transition strength of multipole operators.
\begin{eqnarray}
\hat{Q}_{LK}\equiv
	\sum_{q=n,p}\sum_{s=\pm 1/2}
	\int r^l Y_{LK}(\hat{r})\hat{\psi}_{sq}^\dagger(\vec{r})
	\hat{\psi}_{sq}(\vec{r}) d\vec{r}.
\label{q30}
\end{eqnarray}
The magnetic quantum number $K$ is defined with respect to the
axis of deformation.
Each RPA mode has the eigenfrequency $\omega$ and
the generators $\hat{P}(q=0)$ and $\hat{Q}(q=0)$.
Taking a suitable linear combination of $\hat{Q}_{LK}$ and $\hat{Q}_{L-K}$,
we can make $\hat{Q}_{LK}$ hermitian with real matrix elements.
The transition strengths between the RPA ground state $\ket{0}$ and
excited state $\ket{\omega}$ are calculated as
\begin{eqnarray}
\bra{\omega}\hat{Q}_{LK}\ket{0} =
 \sqrt{\frac{1}{2\omega}} \sum_{n\in p,j\in h}
	(Q_{LK})_{nj} P_{nj} ,
\label{trmx}
\end{eqnarray}
where $(Q_{LK})_{nj}$ are the ph matrix elements of $\hat{Q}_{LK}$.
The NG modes are characterized by the zero energy ($\omega=0$) and by
large matrix elements
of $Q_{LK}$ with $L=1$ (translation) and the $(L,K)=(2,\pm 1)$ (rotation).
In contrast, the reaction path is associated with large transition
strength for $\hat{Q}_{20}$ and/or $\hat{Q}_{30}$.
%for the symmetric reaction system and the octupole operator $\hat{Q}_{30}$
%for the asymmetric system.
%The octupole deformation $Q_{30}$ and quadrupole deformation $Q_{20}$ are
%defined as the the expectation value of operators $\hat{Q}_{30}$ and $\hat{Q}_{20}$.
%To identify the translational motion along the axises,
%the $\hat{D}$ operator can be set to $x, y, z$.
%The rotational modes are selected out by setting $\hat{D}$ to be  $\hat{Q}_{2\pm1}$.
Moving away from the initial state, we choose a set of generators
$(\hat{Q}(q),\hat{P}(q))$,
using a condition that the generators must continuously change.

Next we show how to construct the collective path \cite{WN16}.
Although fully self-consistent solution of the moving HF equation (\ref{chf})
is possible, it is significantly facilitated by adopting an approximation,
$\hat{Q}(q+\delta q)\approx \hat{Q}(q)$.
Since $\hat{Q}(q)$ is a smooth function of $q$,
this is reasonable for a small step size $\delta q$.
Thus, the moving Hamiltonian at $q+\delta q$ is now given by
$\hat{H}_{\rm mv}(q+\delta q)=\hat{H}-\lambda \hat{Q}(q)$.
The Lagrange multiplier $\lambda$ is determined by the constraint on
the step size,
\begin{equation}
%\textrm{Tr}[\rho(q+\delta q) \hat{Q}(q) ]
 \bra{\psi(q+\delta q)} \hat{Q}(q) \ket{\psi(q+\delta q)}
% = \sum_i \bra{\varphi_i(q+\delta q)} \hat{Q}(q) \ket{\varphi_i(q+\delta q)}
 = \delta q .
\label{step_size_constraint}
\end{equation}
In this way, the system moves from $\ket{\psi(q)}$ to $\ket{\psi(q+\delta q)}$,
obtaining a new state $\ket{\psi(q+\delta q)}$ on the collective path.

Solving Eqs. (\ref{ASCC1}) and (\ref{ASCC2}) at $\ket{\psi(q+\delta q)}$,
the generators are updated from
$\hat{Q}(q)$ to $\hat{Q}(q+\delta q)$.
Then, we can construct the next state, $\ket{\psi(q+2 \delta q)}$.
Continuing this iteration, we will obtain a series of states,
$\ket{\psi(q=0)}$,
$\ket{\psi(\delta q)}$,
$\ket{\psi(2\delta q)}$,
$\ket{\psi(3\delta q)},\cdots$, forming a collective path.
%This iteration procedure will be implemented to obtain the collective path in
%the next section.
In this work, we set $\delta q$ within the magnitude of
0.1 fm in Eq. (\ref{step_size_constraint}).
In order to check the validity of the approximation
$\hat{Q}(q+\delta q)=\hat{Q}(q)$ at any $q$,
we perform the imaginary-time evolution for the obtained
$\ket{\psi(q)}$ with $\hat{H}_{\rm mv}(q)=\hat{H}-\lambda \hat{Q}(q)$,
and confirm that the state is almost invariant under the iteration.

We should remark a practical treatment of the NG modes.
In principle, the ASCC guarantees the separation of the NG modes from other
normal modes \cite{MNM00,NMMY16}.
However, in this study, we neglect the curvature term
in Eq. (\ref{ASCC2}).
Thus,
at a non-equilibrium point $\ket{\psi(q)}$ away from the ground state,
they can mix with other physical modes of excitation.
Furthermore, in practice, because of the finite mesh size for
the grid representation of the coordinate space (Sec.~\ref{sec:results}),
the exact translational and rotational symmetries are violated.
This is not a problem, if the system has certain symmetries which
prohibit a mixture of the NG modes with physical modes of interest.
For instance, the collision of two $^{16}$O nuclei is free from
the problem,
because the system keeps the parity and the axial symmetry,
thus, the $K^\pi$ quantum numbers clearly separate the NG modes from
the colliding motion of two nuclei.
In contrast, for the asymmetric reaction of $\alpha+^{16}$O,
the NG mode corresponding to
the translation of the center of mass along the symmetric axis can be mixed
with the $K^{\pi} = 0^{-}$ excitation.
In this case we need to remove the NG components
($\hat{Q}^{\rm NG}(q), \hat{P}^{\rm NG}(q)$)
from the ASCC generators, $\hat{Q}^{\rm cal}(q)$ and $\hat{P}^{\rm cal}(q)$.
Since the NG generators
$(\hat{Q}^{\rm NG}, \hat{P}^{\rm NG})$
are trivially obtained in the translational motion,
the ASCC generators for the reaction are easily corrected as
\begin{eqnarray}
\hat{Q}(q)&=&\hat{Q}^{\rm cal}(q)-\lambda^Q_q\hat{Q}^{\rm NG}
 -\lambda^Q_p\hat{P}^{\rm NG}, \\
\hat{P}(q)&=&\hat{P}^{\rm cal}(q)-\lambda^P_q\hat{Q}^{\rm NG}
 -\lambda^P_p\hat{P}^{\rm NG},
\end{eqnarray}
with
\begin{eqnarray}
\lambda^Q_q&=&-i\bra{\psi(q)} [\hat{Q}^{\rm cal},\hat{P}^{\rm NG} \ket{\psi(q)}
, \\
\lambda^Q_p&=& i\bra{\psi(q)} [\hat{Q}^{\rm cal},\hat{Q}^{\rm NG} \ket{\psi(q)}
, \\
\lambda^P_q&=& -i\bra{\psi(q)} [\hat{P}^{\rm cal},\hat{P}^{\rm NG} \ket{\psi(q)}
, \\
\lambda^P_p&=& i\bra{\psi(q)} [\hat{P}^{\rm cal},\hat{Q}^{\rm NG} \ket{\psi(q)}
,
 \label{remNG}
\end{eqnarray}
which can be derived from the condition,
$[\hat{Q}(q),\hat{Q}^{\rm NG} ] = [\hat{Q}(q),\hat{P}^{\rm NG} ] =
[\hat{P}(q),\hat{Q}^{\rm NG} ] = [\hat{P}(q),\hat{P}^{\rm NG} ] = 0$.
%The same procedure is applied to the generator $\hat{P}(q)$.

%%%%%%%%%%%%%%%%%%%%%%%%%%%%%%%%%%%%%%%%%%%%%%%%%%%%%%%%%%%%%%%%%%%%
\subsection{\label{sec:theoc}Inertial mass for nuclear reaction}

As mentioned in Sec. \ref{subsec:theoA},
to fix the arbitrary scale of $q$,
the inertial mass $M(q)$ with respect to $q$ in Eq. (\ref{ASCC2})
is set to be
1 $\hbar^2$MeV$^{-1}$fm$^{-2}$.
In order to obtain a physical picture of the collective
dynamics, it is convenient to label the collective path by
other coordinates intuitively chosen by ourselves.
For instance, in the asymptotic region where the two colliding nuclei
are well apart, it is natural to
adopt the relative distance $R$ between projectile and target.
As far as the one-to-one correspondence between $q$ and $R$
is guaranteed, we may use the mapping function $R(q)$ to modify the
scale of the coordinate, without losing anything.
For the coordinate $R$, the inertial mass should be transformed as
\begin{eqnarray}
M(R) =M_q\left(\frac{dq}{dR}\right)^{2}
= M_q \left(\frac{dR}{dq}\right)^{-2} .
\label{mass}
\end{eqnarray}
Thus, the mass $M(R)$ requires the calculation of the derivative $dR/dq$,
which can be obtained as%by use of
%the local generator $\hat{P}(q)$,
\begin{eqnarray}
\frac{dR}{dq} &=&
\frac{d}{dq}\langle \psi(q) |\hat{R}| \psi(q) \rangle
 =\langle \psi(q) | [\hat{R},\frac{1}{i}\hat{P}(q) ] | \psi(q) \rangle
 \nonumber \\
% &=&\sum_{mi}\left\{ R_{mi}(q)P_{mi}^*(q)+R_{mi}^*(q)P_{mi}(q)\right\}
 &=&2\sum_{n\in p, j\in h} R_{nj}(q)P_{nj}(q)
,\label{mass3}
\end{eqnarray}
with the local generator $\hat{P}(q)$.
$R_{nj}(q)$ are the ph matrix elements of $\hat{R}$.
%with respect to the state $\ket{\psi(q)}$.
%An alternative way to calculate the finite difference,
%${dR}/{dq}\approx \{R(q+\delta q)-R(q)\}/{\delta q}$,
%with two adjacent points, $\ket{\psi(q)}$ and $\ket{\psi(q+\delta q)}$,

%For the study of nuclear fusion/fission reactions,
In this paper, the one-body operator $\hat{R}$ for the relative distance
between projectile and target is defined as follows.
Assuming the relative motion along z axis with projectile on the right and
target on the left,
we introduce a separation plane at $z=z_{\rm s}$ so that
\begin{eqnarray}
\int_{-\infty}^\infty dx \int_{-\infty}^\infty dy
	\int_{z_{\rm s}}^{+\infty} dz \rho(\vec{r})= A_{\rm pro},
\end{eqnarray}
where $\rho(\vec{r})$ is the total density,
$A_{\rm pro}$ ($A_{\rm tar}$) is the mass number of the projectile (target).
The operator form of $R$ reads
%asymmetric reaction system is defined as
\begin{eqnarray}
	\hat{R}\equiv \sum_{s,q} \int d\vec{r}
	\hat{\psi}_{sq}^\dagger(\vec{r})\hat{\psi}_{sq}(\vec{r})
         z\left[\frac{\theta(z-z_{\rm s})}{A_{\rm pro}}-
         \frac{\theta(z_{\rm s}-z)}{A_{\rm tar}}\right]  ,\label{defr}
\end{eqnarray}
where $\theta$ is the step function.
For the symmetric reaction system, the section plane is at $z_{\rm s} = 0$.
$\hat{R}$ reduces to
\begin{eqnarray}
	\hat{R}=\frac{2}{A} \sum_{s,q}\int d\vec{r} z \hat{\psi}_{sq}^\dagger(\vec{r})\hat{\psi}_{sq}(\vec{r})\left[\theta(z)-\theta(-z)\right].\label{rdef}
\end{eqnarray}

%%%%%%%%%%%%%%%%%%%%%%%%%%%%%%%%%%%%%%%%%%%%%%%%%%%%%%%%%%%%%%%%%%%%
\section{\label{sec:results}Applications}{\label{sec:path}

%\section{\label{sec:results}ASCC collective motion path}{\label{sec:path}
%%%%%%%%%%%%%%%%%%%%%%%%%%%%%%%%%%%%%%%%%%%%%%%%%%%%%%%%%%%%%%%%%%%%

In this section we present results of numerical application
to the fusion reaction.
%Different features and dynamical informations
%depending on the system will be demonstrated.
We employ the Bonche-Koonin-Negele (BKN) EDF
\cite{BKN76}, which assumes the spin-isospin symmetry
without the spin-orbit interaction.
%The parameter set is chosen to be the same as Ref. \cite{BKN76}.
To express the orbital wave functions,
the grid representation is employed, discretizing the rectangular box into
the three-dimensional (3D) Cartesian mesh.
The model space is set to be $12\times12\times18$ fm$^{3}$ for the
%reaction systems of $^{16}$O $\leftrightarrow$ $^{12}$C+$\alpha$
reaction $^{16}$O+$\alpha$ $\rightarrow$ $^{20}$Ne,
$12\times12\times24$ fm$^{3}$ for the system $^{16}$O+$^{16}$O $\rightarrow$ $^{32}$S,
the mesh size is set to be 1.1 fm.

%%%%%%%%%%%%%%%%%%%%%%%%%%%%%%%%%%%%%%%%%%%%%%%%%%%%%%%%%%%%%%%%%%%%
\subsection{$^{16}$O+$\alpha$ $\rightarrow$ $^{20}$Ne}

\subsubsection{\label{sec:asccNe}Collective path: $^{16}$O$+\alpha\rightarrow
$ ground-state $^{20}$Ne}

As a trivial solution of the ASCC equations,
the well separated $^{16}$O and $\alpha$ both at the ground states
can be the initial state $|\psi(q=0)\rangle$
to start the iterative procedure
in Sec. \ref{sec:theob}.
Alternatively the ground state of $^{20}$Ne can also
be the initial state for the iteration.
%Alternatively the iteration may also start from the $^{16}$O + $\alpha$
Although it is not trivial, we find that the same trajectory
is produced starting from these two initial states.
The ASCC collective path smoothly connects the two well separated
nuclei, $^{16}$O and $\alpha$, to $^{20}$Ne at ground state.
The ground state of $^{20}$Ne has a large quadrupole deformation.
The density profile is shown in Fig. \ref{fig:asdens} (d).
At the ground state,
the lowest physical RPA state is found to be the $K^\pi=0^-$
octupole excitation, which
has a sizable transition strength of the operator $\hat{Q}_{30}$
defined in Eq. (\ref{q30}).
Choosing this $K^\pi=0^-$ octupole mode as the generators
$\hat{Q}(q)$ and $\hat{P}(q)$,
a series of states can be obtained by iteration,
forming a collective fusion path of
$^{16}$O+$\alpha$ $\leftrightarrow$ $^{20}$Ne.
In the asymptotic region (Fig.~ \ref{fig:asdens} (a)),
the generators smoothly change into those representing the relative motion
between $^{16}$O and $\alpha$.
Figure \ref{fig:asdens} shows density distributions
in the x-z plane ($y=0$) at four different points
on the collective path.
The panel (a) shows the well separated $^{16}$O + $\alpha$,
the panel (d) shows $^{20}$Ne at ground state,
and two intermediate states are shown in the panels (b) and (c).

\begin{figure}
\begin{centering}
\includegraphics[width=0.90\columnwidth]{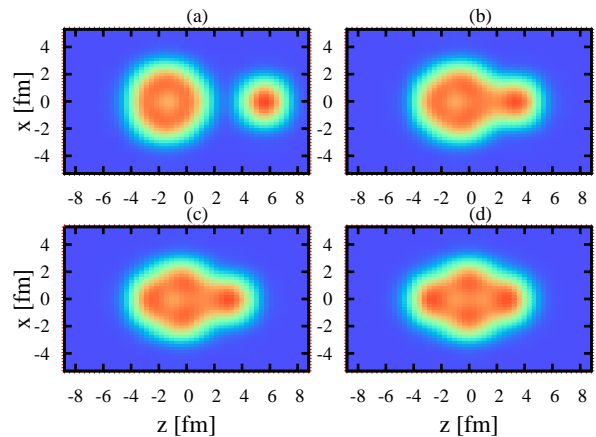}
\par\end{centering}
\caption{\label{fig:asdens}(Color online)
Density distribution on the x-z plane at four points on the ASCC
fusion reaction path of $^{16}$O + $\alpha \rightarrow ^{20}$Ne:
(a) $R=7.6$ fm,
(b) $R = 5.2$ fm,
(c) $R = 4.2$ fm, and
(d) $R=3.8$ fm corresponding to the ground state of $^{20}$Ne.
}
\end{figure}

\begin{figure}
\begin{centering}
\includegraphics[width=0.90\columnwidth]{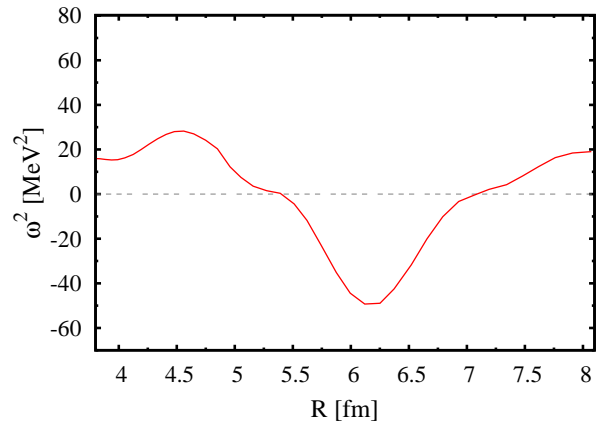}
\par\end{centering}
\caption{\label{fig:neomega}(Color online)
Square of the moving RPA eigenfrequency $\omega^{2}(q)$
on the ASCC collective path
of $^{16}$O + $\alpha \rightarrow ^{20}$Ne,
shown as a function of relative distance $R$.
At the ground state of $^{20}$Ne ($R=3.8$ fm),
this mode corresponds to the $K^{\pi} = 0^{-}$ octupole mode of
excitation.
}
\end{figure}

Figure \ref{fig:neomega} shows the square of moving RPA eigenfrequency
$\omega^{2}(q)$ of the generators with $K = 0$,
as a function of relative distance $R$.
At the ground state of $^{20}$Ne ($R=3.8$ fm),
the parity is a good quantum number and the RPA mode corresponds to
the negative parity $\pi=-$,
leading to $\bra{\omega}\hat{Q}_{30}\ket{0} = 3017$ fm$^3$ and
$\bra{\omega}\hat{Q}_{20}\ket{0} = 0$.
At larger $R$, the octupole deformation $Q_{30}$ increases, then,
the parity is no longer conserved.
The transition strength $\bra{\omega}\hat{Q}_{20}\ket{0}$ becomes
nonzero, then, gradually changes its character into the relative motion
between $^{16}$O and $\alpha$.
%{\bf
%We confirm that the eigenfrequency of the moving RPA
%coincides with the second derivative of the potential, Eq. (\ref{ab4}).
%}
Since the curvature of the potential energy can be negative,
the value of
$\omega^{2}(q)$ can be negative leading to imaginary $\omega(q)$.
Since the generators keep the $K=0$ character all the way,
the states $\ket{\psi(q)}$ on the collective path are axially symmetric.
%Below this lowest physical mode of $K=0$,
There appear five NG modes, namely,
two rotational modes, and the three translational modes.
In actual calculation, these NG modes have finite energy
due to the finite mesh size in numerical calculation.
At the ground state, we obtain
$\omega = 1.9$ MeV for the rotational modes,
$\omega = 3.5$ MeV for the translational modes along x and y directions,
and $\omega = 1.3$ MeV for the translational mode along z direction.
%the same for the other reaction systems in the next two subsections.
%Because of the axial symmetry of the ground state, the rotation
%about the symmetry axis (z axis) does not appear.
%The pink line labeled by $\langle n|Y_{21, 2-1}|0\rangle \ne 0$
%represent the rotational states.
%The ones labeled by $\langle n|X,Y,Z|0\rangle \ne 0$ represent the three
%translation states. Since the system keeps axial symmetry along Z axis on
%the ASCC path, the two translational states in X and Y directions as well as
%the two rotational states are degenerated as show in the figure.

%If the iteration start from the ground state of $^{20}$Ne, %$Q_{30}  = 0$,
%initially the transition strength of $\hat{Q}_{20}$ for this selected $0^{-}$
%octupole RPA state is zero.
%the lowest physical mode in energy,
%which has a sizable transition strength of operator
%$\hat{Q}_{30}$,
%and  . %$\langle n|Y_{20}|0\rangle = 0$.
%At the ground state of $^{20}$Ne,
The next lowest $K=0$ mode of excitation at
the ground state of $^{20}$Ne
has the positive parity $\pi=+$ and
a transition strength of operator of $\hat{Q}_{20}$,
%$\bra{\omega}\hat{Q}_{30}\ket{0} = 0$ fm$^3$ and
$\bra{\omega}\hat{Q}_{20}\ket{0} = 5.3$ fm$^{2}$.
The RPA frequency
$\omega$ of this state is about 10 MeV, which is much higher than the
octupole mode and many other modes with $K\neq 0$.
If we adopt this $K^\pi=0^{+}$ mode as the starting generators,
we cannot construct the collective path connecting the ground state
and two separated nuclei.
Generally speaking, the higher the RPA eigenfrequency is,
the more difficult it is to find a solution of the moving mean-field
equation (\ref{chf}).
%it turns out to be difficult to obtain the solution of the moving meanfield
%equation Eq. (\ref{chf}), where the constraint operator is constructed by
%this high energy state.

%On this collective path that connecting $^{20}$Ne and $^{16}$O + $\alpha$,
%the RPA solution of quadupole mode is always at the same the octupole mode.

\begin{figure}
\begin{centering}
\includegraphics[width=0.90\columnwidth]{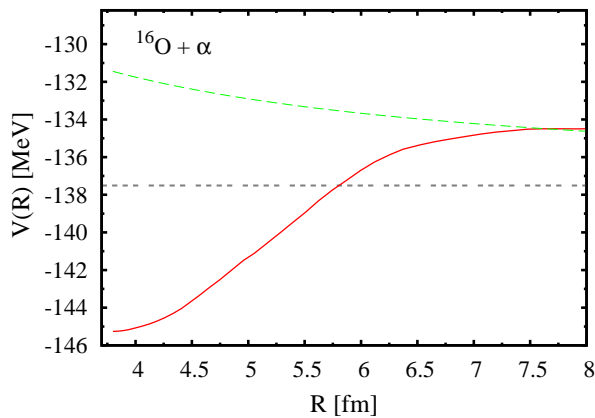}
\par\end{centering}
\caption{\label{fig:nee}(Color online)
Potential energy for the fusion path
$^{16}$O + $\alpha \rightarrow ^{20}$Ne as a function of
relative distance $R$.
The solid (red) line corresponds to $V(R)$ on the ASCC collective
path, while the dashed (green) line shows
$16e^{2}/R+E_{\rm gs}(\alpha)+E_{\rm gs}(^{16}\mbox{\rm O})$ for reference.
The horizontal dashed (grey) line indicates
the asymptotic energy of $E_{\rm gs}(^{16}\mbox{\rm O})+E_{\rm gs}(\alpha)$.
}
\end{figure}
\begin{figure}
\begin{centering}
\includegraphics[width=0.90\columnwidth]{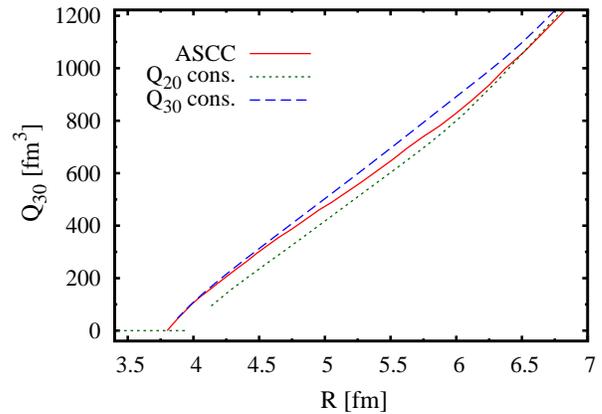}
\par\end{centering}
\caption{\label{fig:y2y3}(Color online)
Octupole $Q_{30}$ moment %(upper panel) and
%quadrupole moments $Q_{20}$ (lower panel)
as a function of relative distance $R$.
The solid (red) line indicates the result of ASCC collective fusion
path of $^{16}$O + $\alpha \rightarrow ^{20}$Ne,
the dotted (green) and dashed (blue) lines indicate
the results of CHF calculation with constraint on
$Q_{20}$ and $Q_{30}$, respectively.
}
\end{figure}

Figure~\ref{fig:nee} shows the potential energy of the ASCC collective path,
Eq. (\ref{vdef}),
as a function of $R$.
The dashed line shows the asymptotic
Coulomb energy on top of the summed ground state energies of
$\alpha$ and $^{16}$O.
The ground state of $^{20}$Ne is at $R = 3.8$ fm, and
the top of the Coulomb barrier is located at $R = 7.7$ fm.
To compare the ASCC collective path with those obtained with
conventional CHF calculations,
we show the octupole %and the quadrupole
moment as a function of $R$ in Fig. \ref{fig:y2y3},
for these different collective paths.
Two collective paths of the CHF calculations
are constructed with the constraining operators
of $\hat{Q}_{20}$ (dotted line) and $\hat{Q}_{30}$ (dashed line).
From Fig. \ref{fig:y2y3} we can see all these three collective paths
deviate from each other.
Especially, for the CHF calculation with quadrupole constraint of
$\hat{Q}_{20}$,
the collective path is not continuous due
to sudden change of the state at around $R = 4$ fm.
%Comparing the ASCC path with the octupole-constrained path,
%the quadrupole deformations $Q_{20}$
%are very close to each other, while the octupole deformations are different.
%compared with the potential energy of the constrained Hartree-Fock(CHF) states
%with different constraints operators $ \hat{Q}_{20}$ and $ \hat{Q}_{30}$.
%At the ground state of $^{20}$Ne or at large distance, the ASCC collective
%path gives the same states as the collective
%path of the CHF states with constraint $\hat{Q}_{30}$ or $\hat{Q}_{20}$.
%where the $^{16}$O and $\alpha$ are well seperated.
%As the two nuclei get closer, the three paths deviate from each other. %of the
%which are shown by the blue and green dash lines respectively.
%For the potential energy of the CHF states with constraint $\hat{Q}_{20}$,
%in the region $0$ fm$^{3} < \hat{Q}_{30} < 200$ fm$^{3}$,
%the potential is not continuous due
%to the change of configuration.
%where the CHF states with constraint $\hat{Q}_{20} $ is give a litter higher
%energy.
%In most of the region, the self-consistent adiabatic collective motion path
%may be regarded approximately as along the constrained HF states,
%this fact is no longer true in the case of the system $^{12}$C + $\alpha$ as
%shown in the next subsection.

%%%%%%%%%%%%%%%%%%%%%%%%%%%%%%%%%%%%%%%%%%%%%%%%%%%%%%%%%%%%%%%%%%%%
\subsubsection{Inertial mass}

%The ASCC method provides us the inertial mass parameter
%with respect to the collective coordinate $q$.
At the Hartree-Fock ground state, the ASCC inertial mass
coincides with the RPA inertial mass
which is able to take into account effect of the time-odd mean fields
\cite{NMMY16}.
Performing the transformation of Eq.~(\ref{mass}),
we may obtain those with respect to the relative distance $R$.
In the asymptotic region of large values of $R$, we expect the inertial
mass becomes identical to the reduced mass of projectile and target,
$\mu_{\rm red}=A_{\rm pro}A_{\rm tar}m/(A_{\rm pro}+A_{\rm tar})$
where $m$ is the nucleon mass.
In most of phenomenological model, in fact,
the mass parameter with respect to $R$ is assumed to be
a constant value of $\mu_{\rm red}$.
In the present microscopic treatment, we may study how the inertial mass
changes during the collision.

One of the most common approaches to the nuclear collective motion is
``CHF+cranking'' approach \cite{Bar11}:
The collective path is produced by the CHF calculation with
a given constraining operator $\hat{O}$, and the inertial mass is
calculated with the cranking formula.
%We use the two types of constraints in this study;
%$\hat{O}=\hat{Q}_{30}$ and $\hat{Q}_{20}$.
Since the quadrupole operator cannot produce a continuous path,
we here use the octupole operator, $\hat{O}=\hat{Q}_{30}$,
to construct the path.
For the cranking mass, we adopt two types of widely-used formulae.
The original formula is derived by the adiabatic perturbation \cite{RS80}.
For the 1D collective path constructed by the CHF calculation with
a given constraining operator $\hat{O}$, it reads
\begin{equation}
M_{\rm cr}^{\rm NP}(R)=2 \sum_{n\in p,j\in h}
\frac{|\bra{\varphi_n(R)}\partial/\partial R\ket{\varphi_j(R)}|^2}
{e_n(R)-e_j(R)} ,
\label{NP_cranking}
\end{equation}
where the single-particle states and energies are defined with
respect to $h_{\rm CHF}(R)=h_{\rm HF}[\rho]-\lambda(R) \hat{O}$ as
\begin{equation}
h_{\rm CHF}(R)\ket{\varphi_\mu(R)}=e_\mu(R))\ket{\varphi_\mu(R)} ,
\quad \mu \in p, h .
\label{CHFe}
\end{equation}
$h_{\rm HF}[\rho]$ is the single-particle mean-field Hamiltonian
reduced from $H$.
%Note that, depending on choice of the constraint operator,
%$\hat{O}=(\hat{Q}_{20},\hat{R})$,
%we obtain slightly different $\ket{\varphi_i(R)}$ even at the same $R$.

Another cranking formula, which is more frequently used in many applications,
is derived, by assuming the separable interaction and taking
the adiabatic limit of the RPA inertial mass,
\begin{equation}
M_{\rm cr}^{\rm P}(R)=
\frac{1}{2} \left\{ S^{(1)}(R)\right\}^{-1}S^{(3)}(R)\left\{S^{(1)}(R)\right\}^{-1} ,
\label{P_cranking}
\end{equation}
with
\begin{equation}
S^{(k)}(R)=\sum_{n\in p,j\in h}\frac{|\bra{\varphi_n(R)}\hat{R}\ket{\varphi_j(R)}|^{2}}
{\{e_{n}(R)-e_{j}(R)\}^{k}}.
\label{S_k}
\end{equation}
According to Ref.~\cite{Bar11},
we call the former one in Eq. (\ref{NP_cranking})
``non-perturbative'' cranking inertia and the latter in Eq. (\ref{P_cranking})
``perturbative'' one.
%The method of CHF + cranking inertia has been widely used for
%many applications, including
%studies of nuclear structure
%\cite{BK68-1,BK68-2,YLQG99,PR09,Nik09,Li09,Del10,Li10,Li11}
%and fission dynamics \cite{WERP02,Bar11,SMBDNS13}.
%To apply the cranking formulae given above,
%the wave function needs to be prepared beforehand
%by the CHF calcultion.
In contrast to the ASCC/RPA mass,
the cranking masses of Eqs. (\ref{NP_cranking}) and (\ref{P_cranking})
both neglect the residual effect.
The cranking formulae produce a wrong total mass for
the translation, when the time-odd mean fields are present.

\begin{figure}
\begin{centering}
\includegraphics[width=0.90\columnwidth]{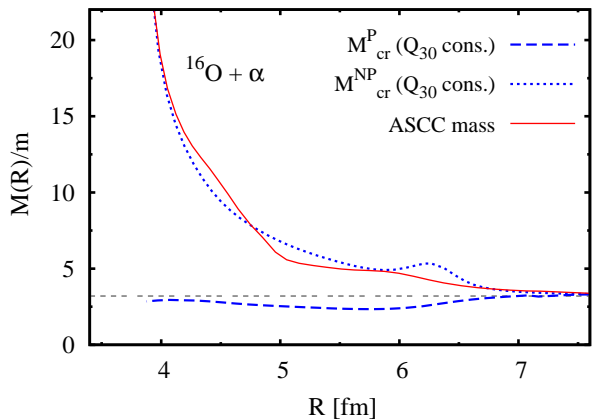}
\par\end{centering}
\caption{\label{fig:massne}(Color online)
The ASCC inertial mass (red solid curve) in units of the nucleon mass
as a function of $R$ for the fusion
path of $^{16}$O + $\alpha \rightarrow ^{20}$Ne, compared
with the cranking inertial masses based on the CHF states
with constraint on $\hat{Q}_{30}$.
%The thin (green) and thick (blue) lines indicate the results
%with constraints on $\hat{Q}_{20}$ and $\hat{Q}_{30}$, respectively.
The non-perturbative and perturbative cranking inertial masses
are shown with dotted and dashed lines, respectively.
}
\end{figure}
Figure \ref{fig:massne} shows the ASCC inertial mass and the cranking masses
for $^{16}$O+$\alpha$ $\rightarrow$ $^{20}$Ne as a function of $R$.
When the two nuclei are far away, the ASCC inertial mass as well as
the cranking masses asymptotically
produce the correct reduced mass of $\mu_{\rm red}=3.2 m$.
%This means at large $R$, the collective coordinate $q$ is parallel
%to the relative distance $R$.
The success of the cranking formulae at large $R$ is
due to the simplicity of the BKN density functional that does not contain
time-odd mean densities.
Thus, this should not be generalized to
more realistic EDFs.
As the projectile and the target approach to each other,
the ASCC inertial mass monotonically increases,
while the cranking masses show different behaviors.
Particularly, the perturbative cranking mass $M^{\rm P}_{\rm cr}(R)$
completely differs from the ASCC and non-perturbative cranking masses.
It is much smaller than the ASCC values and even smaller than
$\mu_{\rm red}$.
The non-perturbative cranking mass based on the $\hat{Q}_{30}$-constrained
path is similar to the ASCC mass.
However, it shows a bump behavior at about $R = 6.3$ fm.
%Since the CHF collective path with constraint on $\hat{Q}_{20}$ is not
%continuous, the cranking masses based on these states
%also show a jump at $R\approx 4$ fm.
%The sudden change of the state does not much affect the
%value of perturbative cranking mass, while
%the non-perturbertive cranking mass has a large discontinuity.

%In addition to these uncertainties
In the cranking formulae,
it is not easy to understand why
the single-particle energies $e_\mu(R)$
in Eqs. (\ref{NP_cranking}) and (\ref{P_cranking})
are defined with respect to $h_{\rm CHF}$ instead of $h_{\rm HF}$.
In contrast, the moving RPA equations (\ref{ASCC1}) and (\ref{ASCC2})
of the ASCC method are invariant with respect to the replacement of
$\hat{H}_{\rm mv}$ with $\hat{H}$.
This is due to the consistency between the constraining operator in
$H_{\rm mv}$ and the generators $\hat{Q}(q)$.
The residual fields induced by the density
fluctuation is properly taken into account in the ASCC mass.

%%%%%%%%%%%%%%%%%%%%%%%%%%%%%%%%%%%%%%%%%%%%%%%%%%%%%%%%%%%%%%%%%%%%
\subsection{\label{sec:resultsc}$^{16}$O+$^{16}$O$\rightarrow^{32}$S
}

\subsubsection{Collective path:
$^{16}\mbox{\rm O}+^{16}\mbox{\rm O}\rightarrow
  \mbox{superdeformed }^{32}\mbox{\rm S}$}

\begin{figure}
\begin{centering}
\includegraphics[width=0.90\columnwidth]{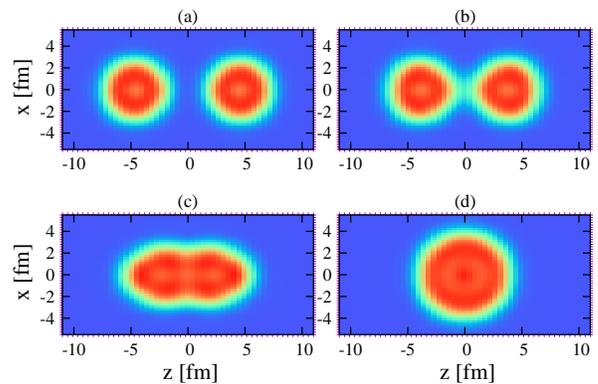}
\par\end{centering}
\caption{\label{fig:snap_shots_OO}(Color online)
Density distribution on the x-z plane at four points on the ASCC
	fusion reaction path of $^{16}$O + $^{16}$O$\rightarrow ^{32}$S:
(a) $R=9.8$ fm,
(b) $R=7.9$ fm corresponding to the barrier top,
(c) $R=4.9$ fm corresponding to the superdeformed $^{32}$S, and
(d) $R=3.7$ fm corresponding to the ground state of $^{32}$S.
}
\end{figure}

\begin{figure}
\begin{centering}
\includegraphics[width=0.90\columnwidth]{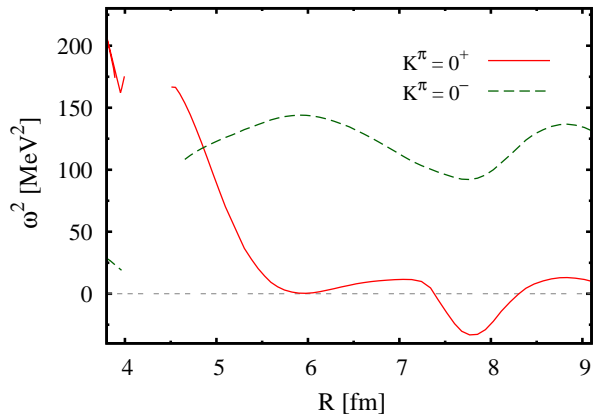}
\par\end{centering}
\caption{\label{fig:ooomega}(Color online)
Square of the RPA eigenfrequency $\omega^{2}(q)$ of the
$K^{\pi} = 0^{+}$ (solid line) and $K^\pi=0^-$ (dashed line) modes
on the ASCC fusion path
of $^{16}$O + $^{16}$O $\rightarrow ^{32}$S,
as a function of relative distance $R$.
Near the ground state of $^{32}$S,
their $K$ quantum numbers are approximate.
}
\end{figure}

We perform the iterative procedure of Sec.~\ref{sec:theob} to
construct the reaction path for $^{16}$O+$^{16}$O.
The initial state of well-separated two $^{16}$O nuclei
is produced by the CHF calculation with
a constraint on the quadrupole moment.
This state corresponds to the separation of $R=9.2$ fm.
The snap shot of the density distribution
is shown in Fig.~\ref{fig:snap_shots_OO} (a).
Figure~\ref{fig:ooomega} shows the value of $\omega^{2}(q)$ of
the  $K^{\pi}=0^{+}$ quadrupole state and the  $K^{\pi}=0^{-}$
octupole state on the ASCC fusion path
as a function of the relative distance $R$.
The local generators $({\hat{P}(q), \hat{Q}(q)})$ of the
$K^\pi=0^{+}$ state is adopted to construct
the collective path.
Except for the three translational and two rotational NG modes,
the eigenfrequency of this $K^\pi=0^{+}$ mode is
the lowest for the region $R > 4.9$ fm.
As the two $^{16}$O approach each other at $R<6$ fm,
the $\omega^{2}(q)$ of the $K^\pi=0^+$ mode
quickly increases and becomes less collective.
At $R < 4.9$ fm, the eigenfrequency of $K^{\pi}=0^{-}$
octupole mode becomes lower than that of the $K^{\pi}=0^{+}$
quadrupole mode.
Energetically favoring the $K^\pi=0^{-}$ mode can be understood
as a tendency to develop an asymmetric shape,
transferring nucleons from one to another.

The obtained potential for $^{16}$O+$^{16}$O
is shown as a function of $R$
in Fig. \ref{fig:ooy20ener}.
For reference,
the dotted line shows the asymptotic
Coulomb energy of $64e^{2}/R+2E_{\rm gs}({^{16}\mbox{O}})$,
where $E_{\rm gs}({^{16}\mbox{O}})$ is the ground state energy of $^{16}$O.
The Coulomb barrier height is about 12.5 MeV
which is located at $R = 7.9$ fm.
Around the barrier top, the curvature of the potential curve is
negative which is consistent with the negative $\omega^2(q)$ in
Fig.~\ref{fig:ooomega}.
Then, the potential reaches a local minimum at $R=4.9$ fm,
which corresponds to the superdeformed (SD) state in $^{32}$S
with $\beta_2=0.94$.
The snap shot of this state is shown in Fig.~\ref{fig:snap_shots_OO} (c).
Beyond the SD state toward even more compact shapes,
the potential shows a significant increase.
In this region, the $K^\pi=0^+$ mode becomes non-collective and
higher in energy,
thus, it is difficult to construct the collective path following
this mode preserving the parity and the axial symmetry.
The ground state of $^{32}$S is located at $R = 3.7$ fm.
We cannot find a self-consistent 1D ASCC path connecting
the SD and the ground states in $^{32}$S.

The ASCC result is compared with that of the
conventional CHF calculation with constraint on $\hat{Q}_{20}$
(dashed line in Fig.~\ref{fig:ooy20ener}).
In the region of $R > 4.9$ fm,
the ASCC collective potential is close to
that of the CHF calculation.
At $R  < 4.9$ fm,
the CHF result deviates from the ASCC potential.
This CHF calculation also produces the local minimum state
at $R = 4.9$ fm,
which confirms that the state reached by
the ASCC path is really the SD minimum.
%At $R = 4.2$ fm, a discontinuous configuration change occurs
%in the CHF calculation.

Figure \ref{fig:spe} shows the single particle energies
of the occupied states of the ASCC path, compared with those
of the CHF path. The CHF single particle energy is defined
in Eq. ({\ref{CHFe}}).
Similarly,
the ASCC single particle energies are defined as
the eigenvalues of $h_{\rm mv}(q)=h_{\rm HF}[\rho(q)]-\lambda(q)\hat{Q}(q)$
%that is the single-particle Hamiltonian reduced from
%$\hat{H}_{\rm mv}(q)=\hat{H}-\lambda(q) \hat{Q}(q)$
with $\lambda(q)=\partial V/\partial q$.
From Fig.~\ref{fig:spe}, we can see the difference between the two set of
single particle energies.
They are identical at the local equilibrium states, namely,
at the ground state of $^{32}$S,
at the SD state at $R = 4.9$ fm, and at large distance $R > 9.0$ fm where the
two $^{16}$O are well separated.
For the CHF calculation, a level crossing at the Fermi level occurred
at $R = 4.3$ fm.
The crossing causes a sudden shape change from
the axially symmetric shapes at large $R$ into
the triaxial shapes at $R < 4.3$ fm. 
This discontinuous configuration change may produce a ``multi-valued'' 
potential as a function of $R$ in Fig.~\ref{fig:ooy20ener}. 
Around the peak of the potential at $R=4.3$ fm, 
the one-to-one correspondence between $Q_{20}$ and $R$ no
longer exist.
%This discontinuous configuration change leads to ``multivalued'' potential
%and single-particle energies as functions of $R$,
%as shown by Fig.~\ref{fig:ooy20ener} and Fig.~\ref{fig:spe}.
In the case of ASCC, the single-particle energies show more moderate
behaviors.
The axial symmetry is kept in the region of $R > 4.5$ fm,
but beyond this region,
we cannot find the ASCC collective path toward more compact shape.
It is not clear yet whether this is due to the level crossing effect
seen in the CHF calculation. Nevertheless, we may speculate that
the ASCC collective path tries to avoids this level crossing,
which leads to the coordinate $q$ almost orthogonal to $R$.
See also discussion on the inertial mass in Sec. \ref{sec:oomass}.

With the BKN functional, the calculated ground state
of $^{32}$S is triaxially deformed
with $\beta_{2} = 0.38$ and $\gamma=36^{\circ}$.
In the triaxial state, the $K$-mixing takes place for the RPA
normal modes.
The lowest physical collective mode at the ground state
is the positive-parity mode with non-zero transition strength
of the operator $\hat{Q}_{22}$.
The RPA mode with a $K^\pi=0^+$ character is located at much
higher in energy.
Following this ``quasi-axial'' mode, we try to construct the
collective path from the ground state ($R=3.7$ fm),
however, we do not succeed to find the ASCC path to connect the SD state
from the ground state.
The path in the region of $4<R<4.5$ fm is still missing in
Figs.~\ref{fig:ooy20ener} and \ref{fig:spe}.
In this region, the triaxial and octupole degrees of freedom may play
an important role, because their frequencies are lower than that
of the quasi-axial ``$K^\pi=0^+$''-like mode.
%This may suggest the necessity to change the generator $\hat{Q}$
%of quadrupole type
%to other excitation modes of moving-RPA on the ASCC path,
%and make a full study of the bifurcations in the
%potential energy surface. This can be achieved
%within the ASCC framework and will be our future study.
%so as to investigate the bifurcations with ASCC.
This may suggest the limitation of the 1D collective path and
necessity to extend to a multi-dimensional collective subspace.
In addition, the pairing effect may change the situation.
It should be also noted that the mixture of the rotational NG modes
due to the missing curvature terms may affect the result
in the triaxial case.

% calculated
%by the procedures introduced in Sec. (\ref{sec:theo}),
%The K = 0, quadrupole excitation mode (red line in Fig. \ref{fig:ooomega})
%is selected to construct the local generator ${\hat{P}(q), \hat{Q}(q)}$.
%We can see that the ASCC collective path does not
%connects two $^{16}$O with the ground state of $^{32}$S,
%In the region of $4.1$ fm $< R < 4.5$ fm
%The ASCC collective path does not
%connects two $^{16}$O with the ground state of $^{32}$S,
%.
%For the ASCC fusion path from $Q_{20}  = 1300$ fm$^{2}$ to
%$Q_{20}  = 320$ fm$^{2}$, the axial symmetry keep wells as the two
%$^{16}$O approach to each other.
%For the region $110$ fm$^{2} < Q_{20}  < 190$ fm$^{2}$,
%where the collective path starts from the ground state of $^{32}$S,
%the system is in the triaxial deformation.
\begin{figure}
\begin{centering}
\includegraphics[width=0.90\columnwidth]{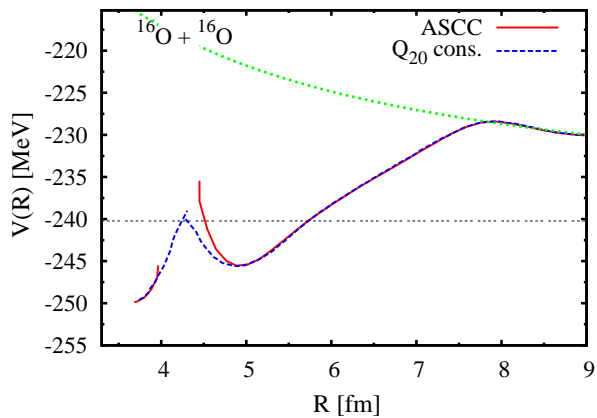}
\par\end{centering}
\caption{\label{fig:ooy20ener}(Color online)
Potential energy of the fusion path
$^{16}$O + $^{16}$O $\rightarrow ^{32}$S as a function of
relative distance $R$.
The solid (red) line indicates the result of ASCC method.
The dashed (blue) line indicates the result of CHF calculation.
The thin dotted (green) line shows
$64e^{2}/R+2E_{\rm gs}(^{16}\mbox{\rm O})$ for reference.
The horizontal dashed (grey) line indicates
the asymptotic energy of $2E_{\rm gs}(^{16}\mbox{\rm O})$.
}
\end{figure}
\begin{figure}
\begin{centering}
\includegraphics[width=0.90\columnwidth]{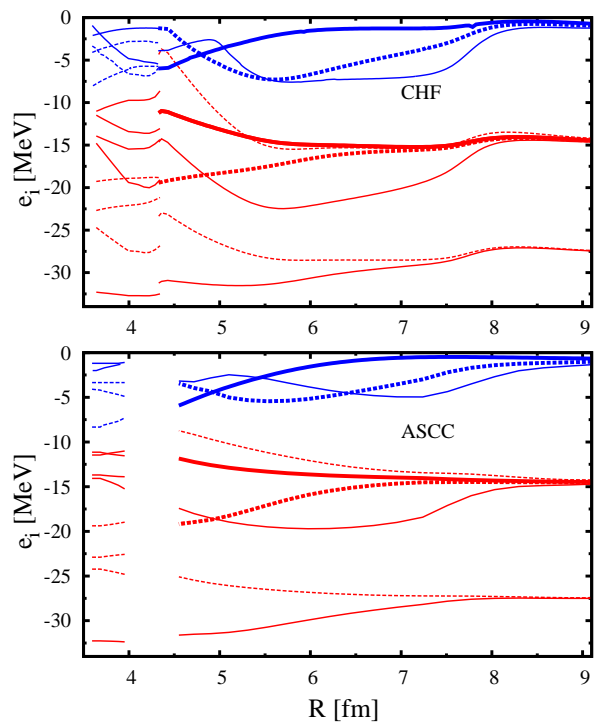}
\par\end{centering}
\caption{\label{fig:spe}(Color online)
Single particle energies for
the fusion path
$^{16}$O + $^{16}$O $\rightarrow ^{32}$S as a function of
relative distance $R$.
The upper panel shows the result of CHF calculation
with constraint on $\hat{Q}_{20}$.
The lower panel shows the single particle energies
of the ASCC collective path.
Because of the spin-isospin symmetry in the BKN functional,
each orbit has four-fold degeneracy.
The thick lines indicate those with eight-fold degeneracy.
The lowest eight orbits (red) are occupied.
The positive- and negetive-parity states are shown by solid and
dashed lines, respectively.
}
\end{figure}

\begin{figure}
\begin{centering}
\includegraphics[width=0.90\columnwidth]{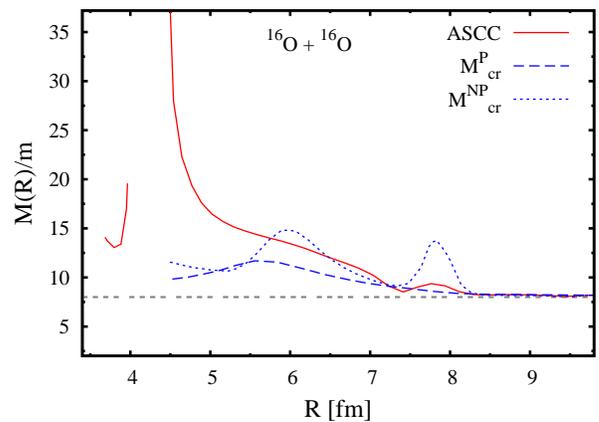}
\par\end{centering}
\caption{\label{fig:massoo}(Color online)
The ASCC inertial mass (red solid curve) in units of the nucleon mass
as a function of $R$ for the fusion
path of $^{16}$O + $^{16}$O $\rightarrow ^{32}$S, compared
with the cranking mass based on the CHF states.
The non-perturbative and perturbative cranking inertial masses
are shown with dotted and dashed lines, respectively.
}
\end{figure}

\subsubsection{\label{sec:oomass}Inertial mass}

Figure~\ref{fig:massoo} shows the inertial mass $M(R)$ for the system
$^{16}$O+$^{16}$O $\rightarrow$ $^{32}$S  as a function of $R$.
For comparison, the perturbative and non-perturbative
cranking masses are also calculated based on the CHF state
with the $\hat{Q}_{20}$ constraint.
The reduced mass, $\mu_{\rm red}=8m$, is well reproduced asymptotically
at large $R$ in both the ASCC and the cranking calculations.
%The broken part around $R = 4.5$ fm corresponds to the
%region $190$ fm$^{2} < Q_{20}  < 320$ fm$^{2}$ in figure \ref{fig:ooy20ener}.
Because of the configuration change of the CHF states,
the cranking masses ($M^{\rm P}_{\rm cr}(R)$ and $M^{\rm NP}_{\rm cr}(R)$)
is discontinuous and jump up to very large values at $R < 4.25$ fm.
They are more than $100 m$, beyond the scale of the vertical axis,
thus, not shown in Fig.~\ref{fig:massoo}.
On the other hand, in the region of $4$ fm $<R<5$ fm,
the ASCC inertial mass $M(R)$ shows a drastic increase 
as decreasing $R$. 
According to Eq. (\ref{mass}),
the large $M(R)$ comes from the large value of $(dR/dq)^{-1}$,
that means
the ASCC reaction path generated by the $K^\pi=0^+$ mode
becomes almost orthogonal to $R$ in the region
between the SD and the ground states in $^{32}$S.

Except for the asymptotic region,
the cranking inertial masses are significantly different
from that of the ASCC.
Furthermore,
the perturbative and the non-perturbative
cranking masses provide different values.
The non-perturbative formula produces oscillating behaviors in
Fig.~\ref{fig:massoo}, which is seen but strongly hindered
in the ASCC calculation.
Since we adopt the BKN density functional
which does not contain time-odd densities,
the different inertial masses
are mainly due to difference in the assumed reaction paths:
In the cranking formula, it is assumed to be the relative distance $R$
between the two $^{16}$O,
while it is the decoupled coordinate $q$
in the ASCC.

%At $R < 5.5$ fm, the ASCC mass increase drastically
%Since the inertial mass is extracted from the quadrupole mode of the collective motion,
%Together with the discussion in Sec. (\ref{sec:theoc}),
%meanwhile it's shown in Sec. (\ref{sec:resultsc}) that  in this region
%the quadrupole motion
%requires larger energy,
%in this case, other collective degree of freedom rather than the symmetric relative
%motion mode is needed or favored to connects the fussion/fission path
%between $^{32}$S and $^{16}$O+$^{16}$O.

%%%%%%%%%%%%%%%%%%%%%%%%%%%%%%%%%%%%%%%%%%%%%%%%%%%%%%%%%%%%%%%%%%%%
\subsubsection{Comparison with former ATDHF calculation}

For the symmetric reaction of $^{16}$O$+^{16}$O,
the result of the ATDHF was reported by Reinhard et al. \cite{RFGGG84}.
The result of Ref.~\cite{RFGGG84} shows the potential $V(R)$
at $R\gtrsim 5$ fm, which look similar to our present result.
Since in their calculation the potential is defined as an envelope of
many ATDHF trajectories,
it is not clear whether the obtained path reaches the
SD local minimum.
Our calculation clearly produces the reaction path connecting
two $^{16}$O nuclei and the superdeformed state in $^{32}$S.

A prominent difference is observed in the calculated inertial masses.
The inertial mass of Ref.~\cite{RFGGG84} resembles
the non-perturbative cranking inertia mass in our calculation,
but differs from the ASCC inertial mass,
especially near the SD state.
Our result shows a peculiar increase in the inertial mass
near the SD local minimum ($R=4.9$ fm).
On the contrary, the ATDHF result of Ref.~\cite{RFGGG84}
even shows a decrease near the ending point at $R\approx 5$ fm.
In our previous study on $\alpha+\alpha\rightarrow^8$Be,
we have also found that the ATDHF potential is relatively similar to
that of the ASCC, while the inertial masses are different.
%We have not understood a reason to cause this difference.

%%%%%%%%%%%%%%%%%%%%%%%%%%%%%%%%%%%%%%%%%%%%%%%%%%%%%%%%%%%%%%%%%%%%
\subsection{Sub-barrier fusion cross section}

The ASCC calculation provides us the collective Hamiltonian
on the optimal reaction path.
Using this, we demonstrate the calculation of sub-barrier fusion cross section
for $^{16}$O+$\alpha$ $\rightarrow$ $^{20}$Ne and
$^{16}$O+$^{16}$O$\rightarrow ^{32}$S.
We follow the procedure in Ref.~\cite{RFGGG84}.
%Because of a schematic nature of the BKN energy density functional,
%we should take this result in a qualitative sense.

Using the collective potential $V(R)$ and the inertial mass $M(R)$
obtained in the ASCC calculation,
the sub-barrier fusion cross section is evaluated with the WKB approximation.
The transmission coefficient for the partial wave $L$
at incident energy $E_{\rm c.m.}$ is given by
\begin{eqnarray}
T_{L}(E_{\rm c.m.}) = [1+\exp(2I_{L})]^{-1},
\end{eqnarray}
with
\begin{eqnarray}
I_{L}(E_{\rm c.m.})&=& \int_{a}^{b}dR \Big\{ 2M(R) \nonumber \\
	&\times& \left( V(R)
   +\frac{L(L+1)}{2\mu_{\rm red}R^{2}} - E_{\rm c.m.} \right) \Big\}^{1/2} ,
	 \label{I_L}
%I_{L}(E_{\rm c.m.})= \int_{a}^{b}\left\{ 2M(R)\left[E-V(R)
% -L(L+1)^{2}/4mR^{2}\right]\right\}^{1/2}dR,
\end{eqnarray}
where $a$ and $b$ are the classical turning points on the inner and outer
sides of the barrier respectively.
The centrifugal potential is approximated as
$L(L+1)/(2\mu_{\rm red} R^{2})$.
%with $\mu$ equal to the reduced mass.
The fusion cross section is given by
\begin{eqnarray}
%\sigma(E_{\rm c.m.}) = \frac{\pi\hbar^{2}}{2\mu E_{\rm c.m.}}\sum_{L}(2L+1)T_{L}(E_{\rm c.m.}).\nonumber \\
	\sigma(E_{\rm c.m.}) = \frac{\pi}{2\mu_{\rm red} E_{\rm c.m.}}\sum_{L}(2L+1)T_{L}(E_{\rm c.m.}) .
	\label{sigma_fus}
\end{eqnarray}
For identical incident nuclei,
Eq. (\ref{sigma_fus}) must be modified according to the proper symmetrization.
Only the partial wave with even $L$ contribute to the cross section as
\begin{eqnarray}
%\sigma(E_{\rm c.m.}) = \frac{\pi\hbar^{2}}{2\mu E_{\rm c.m.}}\sum_{L}[1+(-)^{L}](2L+1)T_{L}(E_{\rm c.m.}).\nonumber \\
	\sigma(E_{\rm c.m.}) = \frac{\pi}{2\mu_{\rm red} E_{\rm c.m.}}\sum_{L}[1+(-)^{L}](2L+1)T_{L}(E_{\rm c.m.}).\nonumber \\
\end{eqnarray}

Instead of $\sigma(E_{\rm c.m.})$, one usually refers to
the astrophysical $S$ factor defined by
\begin{eqnarray}
S(E_{\rm c.m.}) = E_{\rm c.m.}\sigma(E_{\rm c.m.}) \exp[ 2\pi Z_{1}Z_{2}e^{2}/\hbar v ],
\end{eqnarray}
where $v$ is the relative velocity at $R\rightarrow\infty$.
The astrophysical $S$ factor is preferred for sub-barrier fusion
because it removes the change by tens of orders of magnitude
present in the cross section due to the trivial penetration through the Coulomb barrier.
The $S$ factor may reveal in a more transparent way
the influence of the nuclear structure and dynamics.

\begin{figure}
\begin{centering}
\includegraphics[width=0.90\columnwidth]{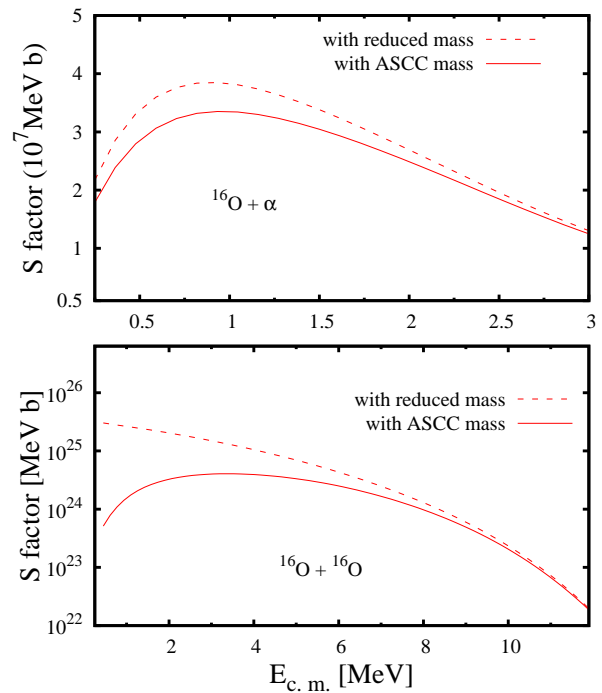}
\par\end{centering}
\caption{\label{fig:sf}(Color online)
The astrophysical $S$ factor for the sub-barrier
fusion of $^{16}$O+$\alpha$ (upper panel) and $^{16}$O+$^{16}$O (lower panel),
as a function of incident energy $E_{\rm c.m.}$.
The solid line indicates the results obtained with the ASCC inertial mass
$M(R)$, the dashed lines
	are calculated with the constant reduced mass $\mu_{\rm red}$.
}
\end{figure}

%\begin{figure}
%\begin{centering}
%\includegraphics[width=0.90\columnwidth]{./SfactorOO.eps}
%\par\end{centering}
%\caption{\label{fig:sfoo}(Color online)
%The astrophysical S factor for the sub-barrier fusion of $^{16}$O+$^{16}$O,
%as a function of incident energy $E_{\rm c.m.}$.
%The solid line indicates the results obtained with the ASCC inertial mass
%$M(R)$, the dashed lines
%are calculated with the constant reduced mass 8$m$.
%}
%\end{figure}

Figure \ref{fig:sf} shows the calculated $S$ factor for the
scattering of $^{16}$O+$\alpha$ and
$^{16}$O+$^{16}$O, respectively.
%For $^{16}$O+$\alpha$, the value of the S factor is plotted in
%normal scale,
For $^{16}$O+$^{16}$O, the values of the $S$ factor are plotted in log scale.
The dashed line is calculated with the same potential $V(R)$
but with the reduced mass,
replacing $M(R)$ by the constant value of $\mu_{\rm red}$ in Eq. (\ref{I_L}).
Effect of the inertial mass is significant in
the deep sub-barrier energy region,
especially for the reaction of $^{16}$O+$^{16}$O at $E_{\rm c.m.} < 4$ MeV.
%For the  result obtained with the coordinate-dependent ASCC inertial mass $M(R)$,
%a bending behavior appears.
Because of a schematic nature of the BKN density functional,
we should regard this result as a qualitative one.
Nevertheless, it suggests the significant effect of the inertial mass
and roughly reproduces
basic features of experimental $S$ factor for the $^{16}$O-$^{16}$O scattering.
This demonstrates the usefulness of the requantization approach
using the ASCC collective Hamiltonian.
%Important quantum effect can be illustrated by the ASCC calculation.

%%%%%%%%%%%%%%%%%%%%%%%%%%%%%%%%%%%%%%%%%%%%%%%%%%%%%%%%%%%%%%%%%%%%

%%%%%%%%%%%%%%%%%%%%%%%%%%%%%%%%%%%%%%%%%%%%%%%%%%%%%%%%%%%%%%%%%%%%
%\section{Inertial mass parameter}{\label{sec:mass}

%%%%%%%%%%%%%%%%%%%%%%%%%%%%%%%%%%%%%%%%%%%%%%%%%%%%%%%%%%%%%%%%%%%%

%%%%%%%%%%%%%%%%%%%%%%%%%%%%%%%%%%%%%%%%%%%%%%%%%%%%%%%%%%%%%%%%%%%%

\section{Summary}\label{sec:sum}

%%%%%%%%%%%%%%%%%%%%%%%%%%%%%%%%%%%%%%%%%%%%%%%%%%%%%%%%%%%%%%%%%%%%

Based on the ASCC method we developed a numerical method to determine the
collective path for the large amplitude nuclear collective motion.
We applied this method to the nuclear fusion reactions;
$^{16}$O+$\alpha\rightarrow^{20}$Ne and
$^{16}$O+$^{16}$O$\rightarrow^{32}$S.
In the grid representation of the 3D coordinate space, the reaction paths,
collective potentials, and the inertial masses are calculated.
%The information of local RPA eigenfrequencies of the collecitve mode on
%the ASCC path are extracted and discussed.

The ASCC collective path smoothly connects the initial state
of $^{16}$O+$\alpha$ to the ground state of the fused nucleus $^{20}$Ne.
%For all the three reaction systems, %of $^{16}$O $\leftrightarrow$ $^{12}$C+$\alpha$
%and $^{32}$S $\leftrightarrow$ $^{16}$O+$^{16}$O,
It is found the self-consistent collective path is different from
that of the conventional CHF calculation with the quadrupole
or octupole moment as the constraint.
For the reaction of $^{16}$O+$^{16}$O$\rightarrow^{32}$S,
we succeed to obtain the 1D reaction path between
$^{16}$O+$^{16}$O and a superdeformed state in $^{32}$S.
The calculated inertial mass asymptotically coincides with the reduced mass,
however, it shows a peculiar increase near equilibrium states,
such as
the ground state of $^{20}$Ne and the superdeformed state of $^{32}$S.

In the present work, we continue to choose the generators of the same
symmetry type, to construct the collective path. In principle we may lift this
restriction.  For instance, inside the superdeformed state of
$^{32}$S, the $K^{\pi}=0^{+}$ quadrupole mode is no longer favored in energy, which
may suggest the necessity to change the generator $\hat{Q}$ of quadrupole type
to octupole type. The importance of the octupole shape in this region was
also suggested in Ref \cite{NJ89}. The bifurcation of the collective
path is possible in the ASCC and will be a future issue.
%In the present work,
%we constantly choose the same excitation of quadrupole type to
%construct the collective coordinate, this quadrupole mode is found
%to be no longer favored in energy inside the superdeformed state of $^{32}$S,
%which suggests the necessity to change the generator $\hat{Q}$
%of quadrupole type
%to other excitation modes of the moving-RPA Eqs. (\ref{ASCC1}-\ref{ASCC2}),
%and make a full study of the bifurcations in the
%potential energy surface. This can be achieved
%within the ASCC framework and will be our future study.

From the ASCC results,
it is straightforward to construct and quantize the
collective Hamiltonian,
to study the collective dynamics microscopically.
The calculated fusion cross section suggests that
the behavior of the inertial mass may have a significant impact on the
fusion probability at deep sub-barrier energies.

Between the superdeformed and triaxial ground states in $^{32}$S,
we cannot find a 1D collective path to connect them.
Since we made an approximation neglecting the curvature terms,
the mixture of the rotational NG modes takes place in the triaxial states.
The multi-dimensional collective subspace may be necessary,
which is beyond the scope of the present work.
%it would be interesting to introduce other collective degree of freedom
%into the ASCC collective path, which would be our future work.
In the present study, the schematic EDF of the BKN
is adopted.
In order to make more quantitative discussion and
apply the method to heavier nuclei, it is
necessary to use realistic EDFs,
and include the pairing correlation.
These are our future tasks.
%By using this method it's feasible to calculated the
%mass parameter with respect to different collective coordinates,

%%%%%%%%%%%%%%%%%%%%%%%%%%%%%%%%%%%%%%%%%%%%%%%%%%%%%%%%%%%%%
% Acknowledgments
%%%%%%%%%%%%%%%%%%%%%%%%%%%%%%%%%%%%%%%%%%%%%%%%%%%%%%%%%%%%%
\begin{acknowledgments}
This work is supported in part by JSPS KAKENHI Grants No. 25287065
and by Interdisciplinary Computational Science Program in CCS,
University of Tsukuba.
%and by ImPACT Program of Council for Science,
%Technology and Innovation (Cabinet Office, Government of Japan).
\end{acknowledgments}

\bibliographystyle{apsrev4-1}
%\bibliographystyle{apalike}
%\begin{thebibliography}{82}%
%\bibliographystyle{apsrev4-1}
%\bibliography{/home/bnlu/Desktop/Works/Documents/Personal.Files/Tetrahedra}
%\bibliography{2012}
\bibliography{./temp,temp11}

\begin{thebibliography}{34}%
\makeatletter
\providecommand \@ifxundefined [1]{%
 \@ifx{#1\undefined}
}%
\providecommand \@ifnum [1]{%
 \ifnum #1\expandafter \@firstoftwo
 \else \expandafter \@secondoftwo
 \fi
}%
\providecommand \@ifx [1]{%
 \ifx #1\expandafter \@firstoftwo
 \else \expandafter \@secondoftwo
 \fi
}%
\providecommand \natexlab [1]{#1}%
\providecommand \enquote  [1]{``#1''}%
\providecommand \bibnamefont  [1]{#1}%
\providecommand \bibfnamefont [1]{#1}%
\providecommand \citenamefont [1]{#1}%
\providecommand \href@noop [0]{\@secondoftwo}%
\providecommand \href [0]{\begingroup \@sanitize@url \@href}%
\providecommand \@href[1]{\@@startlink{#1}\@@href}%
\providecommand \@@href[1]{\endgroup#1\@@endlink}%
\providecommand \@sanitize@url [0]{\catcode `\\12\catcode `\$12\catcode
  `\&12\catcode `\#12\catcode `\^12\catcode `\_12\catcode `\%12\relax}%
\providecommand \@@startlink[1]{}%
\providecommand \@@endlink[0]{}%
\providecommand \url  [0]{\begingroup\@sanitize@url \@url }%
\providecommand \@url [1]{\endgroup\@href {#1}{\urlprefix }}%
\providecommand \urlprefix  [0]{URL }%
\providecommand \Eprint [0]{\href }%
\providecommand \doibase [0]{http://dx.doi.org/}%
\providecommand \selectlanguage [0]{\@gobble}%
\providecommand \bibinfo  [0]{\@secondoftwo}%
\providecommand \bibfield  [0]{\@secondoftwo}%
\providecommand \translation [1]{[#1]}%
\providecommand \BibitemOpen [0]{}%
\providecommand \bibitemStop [0]{}%
\providecommand \bibitemNoStop [0]{.\EOS\space}%
\providecommand \EOS [0]{\spacefactor3000\relax}%
\providecommand \BibitemShut  [1]{\csname bibitem#1\endcsname}%
\let\auto@bib@innerbib\@empty
%</preamble>
\bibitem [{\citenamefont {Schunck}\ and\ \citenamefont {Robledo}(2016)}]{SR16}%
  \BibitemOpen
  \bibfield  {author} {\bibinfo {author} {\bibfnamefont {N.}~\bibnamefont
  {Schunck}}\ and\ \bibinfo {author} {\bibfnamefont {L.~M.}\ \bibnamefont
  {Robledo}},\ }\href {http://stacks.iop.org/0034-4885/79/i=11/a=116301}
  {\bibfield  {journal} {\bibinfo  {journal} {Reports on Progress in Physics}\
  }\textbf {\bibinfo {volume} {79}},\ \bibinfo {pages} {116301} (\bibinfo
  {year} {2016})}\BibitemShut {NoStop}%
\bibitem [{\citenamefont {Brack}\ \emph {et~al.}(1972)\citenamefont {Brack},
  \citenamefont {Damgaard}, \citenamefont {Jensen}, \citenamefont {PauIi},
  \citenamefont {Strutinsky},\ and\ \citenamefont {Wong}}]{Bra72}%
  \BibitemOpen
  \bibfield  {author} {\bibinfo {author} {\bibfnamefont {M.}~\bibnamefont
  {Brack}}, \bibinfo {author} {\bibfnamefont {J.}~\bibnamefont {Damgaard}},
  \bibinfo {author} {\bibfnamefont {A.~S.}\ \bibnamefont {Jensen}}, \bibinfo
  {author} {\bibfnamefont {H.~C.}\ \bibnamefont {PauIi}}, \bibinfo {author}
  {\bibfnamefont {V.~M.}\ \bibnamefont {Strutinsky}}, \ and\ \bibinfo {author}
  {\bibfnamefont {C.~Y.}\ \bibnamefont {Wong}},\ }\href {\doibase
  10.1103/RevModPhys.44.320} {\bibfield  {journal} {\bibinfo  {journal} {Rev.
  Mod. Phys.}\ }\textbf {\bibinfo {volume} {44}},\ \bibinfo {pages} {320}
  (\bibinfo {year} {1972})}\BibitemShut {NoStop}%
\bibitem [{\citenamefont {Brink}\ \emph {et~al.}(1976)\citenamefont {Brink},
  \citenamefont {Giannoni},\ and\ \citenamefont {Veneroni}}]{BGV76}%
  \BibitemOpen
  \bibfield  {author} {\bibinfo {author} {\bibfnamefont {D.~M.}\ \bibnamefont
  {Brink}}, \bibinfo {author} {\bibfnamefont {M.~J.}\ \bibnamefont {Giannoni}},
  \ and\ \bibinfo {author} {\bibfnamefont {M.}~\bibnamefont {Veneroni}},\
  }\href {\doibase 10.1016/0375-9474(76)90004-X} {\bibfield  {journal}
  {\bibinfo  {journal} {Nuclear Physics A}\ }\textbf {\bibinfo {volume}
  {258}},\ \bibinfo {pages} {237} (\bibinfo {year} {1976})}\BibitemShut
  {NoStop}%
\bibitem [{\citenamefont {Villars}(1977)}]{Vil77}%
  \BibitemOpen
  \bibfield  {author} {\bibinfo {author} {\bibfnamefont {F.}~\bibnamefont
  {Villars}},\ }\href {\doibase 10.1016/0375-9474(77)90253-6} {\bibfield
  {journal} {\bibinfo  {journal} {Nucl. Phys. A}\ }\textbf {\bibinfo {volume}
  {258}},\ \bibinfo {pages} {269} (\bibinfo {year} {1977})}\BibitemShut
  {NoStop}%
\bibitem [{\citenamefont {Baranger}\ and\ \citenamefont
  {V{\'{e}}n{\'{e}}roni}(1978)}]{BV78}%
  \BibitemOpen
  \bibfield  {author} {\bibinfo {author} {\bibfnamefont {M.}~\bibnamefont
  {Baranger}}\ and\ \bibinfo {author} {\bibfnamefont {M.}~\bibnamefont
  {V{\'{e}}n{\'{e}}roni}},\ }\href
  {http://www.sciencedirect.com/science/article/pii/0003491678902658}
  {\bibfield  {journal} {\bibinfo  {journal} {Annals of Physics}\ }\textbf
  {\bibinfo {volume} {114}},\ \bibinfo {pages} {123} (\bibinfo {year}
  {1978})}\BibitemShut {NoStop}%
\bibitem [{\citenamefont {Goeke}\ and\ \citenamefont {Reinhard}(1978)}]{GR78}%
  \BibitemOpen
  \bibfield  {author} {\bibinfo {author} {\bibfnamefont {K.}~\bibnamefont
  {Goeke}}\ and\ \bibinfo {author} {\bibfnamefont {P.-G.}\ \bibnamefont
  {Reinhard}},\ }\href
  {http://www.sciencedirect.com/science/article/pii/S0003491678800037}
  {\bibfield  {journal} {\bibinfo  {journal} {Annals of Physics}\ }\textbf
  {\bibinfo {volume} {112}},\ \bibinfo {pages} {328} (\bibinfo {year}
  {1978})}\BibitemShut {NoStop}%
\bibitem [{\citenamefont {Goeke}\ \emph {et~al.}(1983)\citenamefont {Goeke},
  \citenamefont {Gr{\"{u}}mmer},\ and\ \citenamefont {Reinhard}}]{GGR83}%
  \BibitemOpen
  \bibfield  {author} {\bibinfo {author} {\bibfnamefont {K.}~\bibnamefont
  {Goeke}}, \bibinfo {author} {\bibfnamefont {F.}~\bibnamefont
  {Gr{\"{u}}mmer}}, \ and\ \bibinfo {author} {\bibfnamefont {P.-G.}\
  \bibnamefont {Reinhard}},\ }\href
  {http://www.sciencedirect.com/science/article/pii/0003491683900258}
  {\bibfield  {journal} {\bibinfo  {journal} {Annals of Physics}\ }\textbf
  {\bibinfo {volume} {150}},\ \bibinfo {pages} {504} (\bibinfo {year}
  {1983})}\BibitemShut {NoStop}%
\bibitem [{\citenamefont {Reinhard}\ and\ \citenamefont {Goeke}(1987)}]{RG87}%
  \BibitemOpen
  \bibfield  {author} {\bibinfo {author} {\bibfnamefont {P.~G.}\ \bibnamefont
  {Reinhard}}\ and\ \bibinfo {author} {\bibfnamefont {K.}~\bibnamefont
  {Goeke}},\ }\href {http://stacks.iop.org/0034-4885/50/i=1/a=001} {\bibfield
  {journal} {\bibinfo  {journal} {Reports on Progress in Physics}\ }\textbf
  {\bibinfo {volume} {50}},\ \bibinfo {pages} {1} (\bibinfo {year}
  {1987})}\BibitemShut {NoStop}%
\bibitem [{\citenamefont {Mukherjee}\ and\ \citenamefont {Pal}(1982)}]{MP82}%
  \BibitemOpen
  \bibfield  {author} {\bibinfo {author} {\bibfnamefont {A.}~\bibnamefont
  {Mukherjee}}\ and\ \bibinfo {author} {\bibfnamefont {M.}~\bibnamefont
  {Pal}},\ }\href
  {http://www.sciencedirect.com/science/article/pii/037594748290152X}
  {\bibfield  {journal} {\bibinfo  {journal} {Nuclear Physics A}\ }\textbf
  {\bibinfo {volume} {373}},\ \bibinfo {pages} {289} (\bibinfo {year}
  {1982})}\BibitemShut {NoStop}%
\bibitem [{\citenamefont {Dang}\ \emph {et~al.}(2000)\citenamefont {Dang},
  \citenamefont {Klein},\ and\ \citenamefont {Walet}}]{DKW00}%
  \BibitemOpen
  \bibfield  {author} {\bibinfo {author} {\bibfnamefont {G.~D.}\ \bibnamefont
  {Dang}}, \bibinfo {author} {\bibfnamefont {A.}~\bibnamefont {Klein}}, \ and\
  \bibinfo {author} {\bibfnamefont {N.~R.}\ \bibnamefont {Walet}},\ }\href
  {http://www.sciencedirect.com/science/article/pii/S0370157399001192}
  {\bibfield  {journal} {\bibinfo  {journal} {Physics Reports}\ }\textbf
  {\bibinfo {volume} {335}},\ \bibinfo {pages} {93} (\bibinfo {year}
  {2000})}\BibitemShut {NoStop}%
\bibitem [{\citenamefont {Matsuo}\ \emph {et~al.}(2000)\citenamefont {Matsuo},
  \citenamefont {Nakatsukasa},\ and\ \citenamefont {Matsuyanagi}}]{MNM00}%
  \BibitemOpen
  \bibfield  {author} {\bibinfo {author} {\bibfnamefont {M.}~\bibnamefont
  {Matsuo}}, \bibinfo {author} {\bibfnamefont {T.}~\bibnamefont {Nakatsukasa}},
  \ and\ \bibinfo {author} {\bibfnamefont {K.}~\bibnamefont {Matsuyanagi}},\
  }\href {\doibase 10.1143/PTP.103.959} {\bibfield  {journal} {\bibinfo
  {journal} {Prog. Theor. Phys.}\ }\textbf {\bibinfo {volume} {103 (5)}},\
  \bibinfo {pages} {959} (\bibinfo {year} {2000})}\BibitemShut {NoStop}%
\bibitem [{\citenamefont {Nakatsukasa}\ \emph {et~al.}(2016)\citenamefont
  {Nakatsukasa}, \citenamefont {Matsuyanagi}, \citenamefont {Matsuo},\ and\
  \citenamefont {Yabana}}]{NMMY16}%
  \BibitemOpen
  \bibfield  {author} {\bibinfo {author} {\bibfnamefont {T.}~\bibnamefont
  {Nakatsukasa}}, \bibinfo {author} {\bibfnamefont {K.}~\bibnamefont
  {Matsuyanagi}}, \bibinfo {author} {\bibfnamefont {M.}~\bibnamefont {Matsuo}},
  \ and\ \bibinfo {author} {\bibfnamefont {K.}~\bibnamefont {Yabana}},\ }\href
  {\doibase 10.1103/RevModPhys.88.045004} {\bibfield  {journal} {\bibinfo
  {journal} {Rev. Mod. Phys.}\ }\textbf {\bibinfo {volume} {88}},\ \bibinfo
  {pages} {045004} (\bibinfo {year} {2016})}\BibitemShut {NoStop}%
\bibitem [{\citenamefont {Hinohara}\ \emph {et~al.}(2008)\citenamefont
  {Hinohara}, \citenamefont {Nakatsukasa}, \citenamefont {Matsuo},\ and\
  \citenamefont {Matsuyanagi}}]{HNMM08}%
  \BibitemOpen
  \bibfield  {author} {\bibinfo {author} {\bibfnamefont {N.}~\bibnamefont
  {Hinohara}}, \bibinfo {author} {\bibfnamefont {T.}~\bibnamefont
  {Nakatsukasa}}, \bibinfo {author} {\bibfnamefont {M.}~\bibnamefont {Matsuo}},
  \ and\ \bibinfo {author} {\bibfnamefont {K.}~\bibnamefont {Matsuyanagi}},\
  }\href {\doibase 10.1143/PTP.119.59} {\bibfield  {journal} {\bibinfo
  {journal} {Prog. Theor. Phys.}\ }\textbf {\bibinfo {volume} {119}},\ \bibinfo
  {pages} {59} (\bibinfo {year} {2008})}\BibitemShut {NoStop}%
\bibitem [{\citenamefont {Hinohara}\ \emph {et~al.}(2009)\citenamefont
  {Hinohara}, \citenamefont {Nakatsukasa}, \citenamefont {Matsuo},\ and\
  \citenamefont {Matsuyanagi}}]{HNMM09}%
  \BibitemOpen
  \bibfield  {author} {\bibinfo {author} {\bibfnamefont {N.}~\bibnamefont
  {Hinohara}}, \bibinfo {author} {\bibfnamefont {T.}~\bibnamefont
  {Nakatsukasa}}, \bibinfo {author} {\bibfnamefont {M.}~\bibnamefont {Matsuo}},
  \ and\ \bibinfo {author} {\bibfnamefont {K.}~\bibnamefont {Matsuyanagi}},\
  }\href {\doibase 10.1103/PhysRevC.80.014305} {\bibfield  {journal} {\bibinfo
  {journal} {Phys. Rev. C}\ }\textbf {\bibinfo {volume} {80}},\ \bibinfo
  {pages} {014305} (\bibinfo {year} {2009})}\BibitemShut {NoStop}%
\bibitem [{\citenamefont {Hinohara}\ \emph {et~al.}(2010)\citenamefont
  {Hinohara}, \citenamefont {Sato}, \citenamefont {Nakatsukasa}, \citenamefont
  {Matsuo},\ and\ \citenamefont {Matsuyanagi}}]{HSNMM10}%
  \BibitemOpen
  \bibfield  {author} {\bibinfo {author} {\bibfnamefont {N.}~\bibnamefont
  {Hinohara}}, \bibinfo {author} {\bibfnamefont {K.}~\bibnamefont {Sato}},
  \bibinfo {author} {\bibfnamefont {T.}~\bibnamefont {Nakatsukasa}}, \bibinfo
  {author} {\bibfnamefont {M.}~\bibnamefont {Matsuo}}, \ and\ \bibinfo {author}
  {\bibfnamefont {K.}~\bibnamefont {Matsuyanagi}},\ }\href {\doibase
  10.1103/PhysRevC.82.064313} {\bibfield  {journal} {\bibinfo  {journal} {Phys.
  Rev. C}\ }\textbf {\bibinfo {volume} {82}},\ \bibinfo {pages} {064313}
  (\bibinfo {year} {2010})}\BibitemShut {NoStop}%
\bibitem [{\citenamefont {Hinohara}\ \emph {et~al.}(2011)\citenamefont
  {Hinohara}, \citenamefont {Sato}, \citenamefont {Yoshida}, \citenamefont
  {Nakatsukasa}, \citenamefont {Matsuo},\ and\ \citenamefont
  {Matsuyanagi}}]{HSYNMM11}%
  \BibitemOpen
  \bibfield  {author} {\bibinfo {author} {\bibfnamefont {N.}~\bibnamefont
  {Hinohara}}, \bibinfo {author} {\bibfnamefont {K.}~\bibnamefont {Sato}},
  \bibinfo {author} {\bibfnamefont {K.}~\bibnamefont {Yoshida}}, \bibinfo
  {author} {\bibfnamefont {T.}~\bibnamefont {Nakatsukasa}}, \bibinfo {author}
  {\bibfnamefont {M.}~\bibnamefont {Matsuo}}, \ and\ \bibinfo {author}
  {\bibfnamefont {K.}~\bibnamefont {Matsuyanagi}},\ }\href {\doibase
  10.1103/PhysRevC.84.061302} {\bibfield  {journal} {\bibinfo  {journal} {Phys.
  Rev. C}\ }\textbf {\bibinfo {volume} {84}},\ \bibinfo {pages} {061302}
  (\bibinfo {year} {2011})}\BibitemShut {NoStop}%
\bibitem [{\citenamefont {Hinohara}\ \emph {et~al.}(2012)\citenamefont
  {Hinohara}, \citenamefont {Li}, \citenamefont {Nakatsukasa}, \citenamefont
  {Nik\ifmmode \check{s}\else \v{s}\fi{}i\ifmmode~\acute{c}\else \'{c}\fi{}},\
  and\ \citenamefont {Vretenar}}]{HLNNV12}%
  \BibitemOpen
  \bibfield  {author} {\bibinfo {author} {\bibfnamefont {N.}~\bibnamefont
  {Hinohara}}, \bibinfo {author} {\bibfnamefont {Z.~P.}\ \bibnamefont {Li}},
  \bibinfo {author} {\bibfnamefont {T.}~\bibnamefont {Nakatsukasa}}, \bibinfo
  {author} {\bibfnamefont {T.}~\bibnamefont {Nik\ifmmode \check{s}\else
  \v{s}\fi{}i\ifmmode~\acute{c}\else \'{c}\fi{}}}, \ and\ \bibinfo {author}
  {\bibfnamefont {D.}~\bibnamefont {Vretenar}},\ }\href {\doibase
  10.1103/PhysRevC.85.024323} {\bibfield  {journal} {\bibinfo  {journal} {Phys.
  Rev. C}\ }\textbf {\bibinfo {volume} {85}},\ \bibinfo {pages} {024323}
  (\bibinfo {year} {2012})}\BibitemShut {NoStop}%
\bibitem [{\citenamefont {Sato}\ \emph {et~al.}(2012)\citenamefont {Sato},
  \citenamefont {Hinohara}, \citenamefont {Yoshida}, \citenamefont
  {Nakatsukasa}, \citenamefont {Matsuo},\ and\ \citenamefont
  {Matsuyanagi}}]{SHYNMM12}%
  \BibitemOpen
  \bibfield  {author} {\bibinfo {author} {\bibfnamefont {K.}~\bibnamefont
  {Sato}}, \bibinfo {author} {\bibfnamefont {N.}~\bibnamefont {Hinohara}},
  \bibinfo {author} {\bibfnamefont {K.}~\bibnamefont {Yoshida}}, \bibinfo
  {author} {\bibfnamefont {T.}~\bibnamefont {Nakatsukasa}}, \bibinfo {author}
  {\bibfnamefont {M.}~\bibnamefont {Matsuo}}, \ and\ \bibinfo {author}
  {\bibfnamefont {K.}~\bibnamefont {Matsuyanagi}},\ }\href {\doibase
  10.1103/PhysRevC.86.024316} {\bibfield  {journal} {\bibinfo  {journal} {Phys.
  Rev. C}\ }\textbf {\bibinfo {volume} {86}},\ \bibinfo {pages} {024316}
  (\bibinfo {year} {2012})}\BibitemShut {NoStop}%
\bibitem [{\citenamefont {Matsuyanagi}\ \emph {et~al.}(2016)\citenamefont
  {Matsuyanagi}, \citenamefont {Matsuo}, \citenamefont {Nakatsukasa},
  \citenamefont {Yoshida}, \citenamefont {Hinohara},\ and\ \citenamefont
  {Sato}}]{MMNYHS16}%
  \BibitemOpen
  \bibfield  {author} {\bibinfo {author} {\bibfnamefont {K.}~\bibnamefont
  {Matsuyanagi}}, \bibinfo {author} {\bibfnamefont {M.}~\bibnamefont {Matsuo}},
  \bibinfo {author} {\bibfnamefont {T.}~\bibnamefont {Nakatsukasa}}, \bibinfo
  {author} {\bibfnamefont {K.}~\bibnamefont {Yoshida}}, \bibinfo {author}
  {\bibfnamefont {N.}~\bibnamefont {Hinohara}}, \ and\ \bibinfo {author}
  {\bibfnamefont {K.}~\bibnamefont {Sato}},\ }\href
  {http://stacks.iop.org/0954-3899/43/i=2/a=024006} {\bibfield  {journal}
  {\bibinfo  {journal} {Journal of Physics G: Nuclear and Particle Physics}\
  }\textbf {\bibinfo {volume} {43}},\ \bibinfo {pages} {024006} (\bibinfo
  {year} {2016})}\BibitemShut {NoStop}%
\bibitem [{\citenamefont {Nakatsukasa}(2012)}]{Nak12}%
  \BibitemOpen
  \bibfield  {author} {\bibinfo {author} {\bibfnamefont {T.}~\bibnamefont
  {Nakatsukasa}},\ }\href {\doibase 10.1093/ptep/pts016} {\bibfield  {journal}
  {\bibinfo  {journal} {Prog. Theor. Exp. Phys.}\ ,\ \bibinfo {pages} {01A207}}
  (\bibinfo {year} {2012})}\BibitemShut {NoStop}%
\bibitem [{\citenamefont {Wen}\ and\ \citenamefont {Nakatsukasa}(2016)}]{WN16}%
  \BibitemOpen
  \bibfield  {author} {\bibinfo {author} {\bibfnamefont {K.}~\bibnamefont
  {Wen}}\ and\ \bibinfo {author} {\bibfnamefont {T.}~\bibnamefont
  {Nakatsukasa}},\ }\href {\doibase 10.1103/PhysRevC.94.054618} {\bibfield
  {journal} {\bibinfo  {journal} {Phys. Rev. C}\ }\textbf {\bibinfo {volume}
  {94}},\ \bibinfo {pages} {054618} (\bibinfo {year} {2016})}\BibitemShut
  {NoStop}%
\bibitem [{\citenamefont {Davies}\ \emph {et~al.}(1980)\citenamefont {Davies},
  \citenamefont {Flocard}, \citenamefont {Krieger},\ and\ \citenamefont
  {Weiss}}]{DFKW80}%
  \BibitemOpen
  \bibfield  {author} {\bibinfo {author} {\bibfnamefont {K.}~\bibnamefont
  {Davies}}, \bibinfo {author} {\bibfnamefont {H.}~\bibnamefont {Flocard}},
  \bibinfo {author} {\bibfnamefont {S.}~\bibnamefont {Krieger}}, \ and\
  \bibinfo {author} {\bibfnamefont {M.}~\bibnamefont {Weiss}},\ }\href
  {http://www.sciencedirect.com/science/article/pii/0375947480905096}
  {\bibfield  {journal} {\bibinfo  {journal} {Nuclear Physics A}\ }\textbf
  {\bibinfo {volume} {342}},\ \bibinfo {pages} {111} (\bibinfo {year}
  {1980})}\BibitemShut {NoStop}%
\bibitem [{\citenamefont {Nakatsukasa}\ \emph
  {et~al.}(2007{\natexlab{a}})\citenamefont {Nakatsukasa}, \citenamefont
  {Inakura},\ and\ \citenamefont {Yabana}}]{NIY07}%
  \BibitemOpen
  \bibfield  {author} {\bibinfo {author} {\bibfnamefont {T.}~\bibnamefont
  {Nakatsukasa}}, \bibinfo {author} {\bibfnamefont {T.}~\bibnamefont
  {Inakura}}, \ and\ \bibinfo {author} {\bibfnamefont {K.}~\bibnamefont
  {Yabana}},\ }\href {\doibase 10.1103/PhysRevC.76.024318} {\bibfield
  {journal} {\bibinfo  {journal} {Phys. Rev. C}\ }\textbf {\bibinfo {volume}
  {76}},\ \bibinfo {pages} {024318} (\bibinfo {year}
  {2007}{\natexlab{a}})}\BibitemShut {NoStop}%
\bibitem [{\citenamefont {Avogadro}\ and\ \citenamefont
  {Nakatsukasa}(2011{\natexlab{a}})}]{AN11}%
  \BibitemOpen
  \bibfield  {author} {\bibinfo {author} {\bibfnamefont {P.}~\bibnamefont
  {Avogadro}}\ and\ \bibinfo {author} {\bibfnamefont {T.}~\bibnamefont
  {Nakatsukasa}},\ }\href {\doibase 10.1103/PhysRevC.84.014314} {\bibfield
  {journal} {\bibinfo  {journal} {Phys. Rev. C}\ }\textbf {\bibinfo {volume}
  {84}},\ \bibinfo {pages} {014314} (\bibinfo {year}
  {2011}{\natexlab{a}})}\BibitemShut {NoStop}%
\bibitem [{\citenamefont {Avogadro}\ and\ \citenamefont
  {Nakatsukasa}(2013{\natexlab{a}})}]{AN13}%
  \BibitemOpen
  \bibfield  {author} {\bibinfo {author} {\bibfnamefont {P.}~\bibnamefont
  {Avogadro}}\ and\ \bibinfo {author} {\bibfnamefont {T.}~\bibnamefont
  {Nakatsukasa}},\ }\href {\doibase 10.1103/PhysRevC.87.014331} {\bibfield
  {journal} {\bibinfo  {journal} {Phys. Rev. C}\ }\textbf {\bibinfo {volume}
  {87}},\ \bibinfo {pages} {014331} (\bibinfo {year}
  {2013}{\natexlab{a}})}\BibitemShut {NoStop}%
\bibitem [{\citenamefont {Ring}\ and\ \citenamefont {Schuck}(1980)}]{RS80}%
  \BibitemOpen
  \bibfield  {author} {\bibinfo {author} {\bibfnamefont {P.}~\bibnamefont
  {Ring}}\ and\ \bibinfo {author} {\bibfnamefont {P.}~\bibnamefont {Schuck}},\
  }\href@noop {} {\emph {\bibinfo {title} {The Nuclear Many-Body Problem}}}\
  (\bibinfo  {publisher} {Springer-Verlag, New York},\ \bibinfo {year}
  {1980})\BibitemShut {NoStop}%
\bibitem [{\citenamefont {Marumori}\ \emph {et~al.}(1980)\citenamefont
  {Marumori}, \citenamefont {Maskawa}, \citenamefont {Sakata},\ and\
  \citenamefont {Kuriyama}}]{MMSK80}%
  \BibitemOpen
  \bibfield  {author} {\bibinfo {author} {\bibfnamefont {T.}~\bibnamefont
  {Marumori}}, \bibinfo {author} {\bibfnamefont {T.}~\bibnamefont {Maskawa}},
  \bibinfo {author} {\bibfnamefont {F.}~\bibnamefont {Sakata}}, \ and\ \bibinfo
  {author} {\bibfnamefont {A.}~\bibnamefont {Kuriyama}},\ }\href {\doibase
  10.1143/PTP.64.1294} {\bibfield  {journal} {\bibinfo  {journal} {Prog. Theor.
  Phys.}\ }\textbf {\bibinfo {volume} {64}},\ \bibinfo {pages} {1294} (\bibinfo
  {year} {1980})}\BibitemShut {NoStop}%
\bibitem [{\citenamefont {Nakatsukasa}\ \emph
  {et~al.}(2007{\natexlab{b}})\citenamefont {Nakatsukasa}, \citenamefont
  {Inakura},\ and\ \citenamefont {Yabana}}]{Nakatsukasa2007_PRC76-024318}%
  \BibitemOpen
  \bibfield  {author} {\bibinfo {author} {\bibfnamefont {T.}~\bibnamefont
  {Nakatsukasa}}, \bibinfo {author} {\bibfnamefont {T.}~\bibnamefont
  {Inakura}}, \ and\ \bibinfo {author} {\bibfnamefont {K.}~\bibnamefont
  {Yabana}},\ }\href {\doibase 10.1103/PhysRevC.76.024318} {\bibfield
  {journal} {\bibinfo  {journal} {Phys. Rev. C}\ }\textbf {\bibinfo {volume}
  {76}},\ \bibinfo {pages} {024318} (\bibinfo {year}
  {2007}{\natexlab{b}})}\BibitemShut {NoStop}%
\bibitem [{\citenamefont {Avogadro}\ and\ \citenamefont
  {Nakatsukasa}(2011{\natexlab{b}})}]{Avogadro2011_PRC84-214314}%
  \BibitemOpen
  \bibfield  {author} {\bibinfo {author} {\bibfnamefont {P.}~\bibnamefont
  {Avogadro}}\ and\ \bibinfo {author} {\bibfnamefont {T.}~\bibnamefont
  {Nakatsukasa}},\ }\href {\doibase 10.1103/PhysRevC.84.014314} {\bibfield
  {journal} {\bibinfo  {journal} {Phys. Rev. C}\ }\textbf {\bibinfo {volume}
  {84}},\ \bibinfo {pages} {014314} (\bibinfo {year}
  {2011}{\natexlab{b}})}\BibitemShut {NoStop}%
\bibitem [{\citenamefont {Avogadro}\ and\ \citenamefont
  {Nakatsukasa}(2013{\natexlab{b}})}]{Avogadro2013_PRC87-014331}%
  \BibitemOpen
  \bibfield  {author} {\bibinfo {author} {\bibfnamefont {P.}~\bibnamefont
  {Avogadro}}\ and\ \bibinfo {author} {\bibfnamefont {T.}~\bibnamefont
  {Nakatsukasa}},\ }\href {\doibase 10.1103/PhysRevC.87.014331} {\bibfield
  {journal} {\bibinfo  {journal} {Phys. Rev. C}\ }\textbf {\bibinfo {volume}
  {87}},\ \bibinfo {pages} {014331} (\bibinfo {year}
  {2013}{\natexlab{b}})}\BibitemShut {NoStop}%
\bibitem [{\citenamefont {Bonche}\ \emph {et~al.}(1976)\citenamefont {Bonche},
  \citenamefont {Koonin},\ and\ \citenamefont {Negele}}]{BKN76}%
  \BibitemOpen
  \bibfield  {author} {\bibinfo {author} {\bibfnamefont {P.}~\bibnamefont
  {Bonche}}, \bibinfo {author} {\bibfnamefont {S.}~\bibnamefont {Koonin}}, \
  and\ \bibinfo {author} {\bibfnamefont {J.~W.}\ \bibnamefont {Negele}},\
  }\href {\doibase 10.1103/PhysRevC.13.1226} {\bibfield  {journal} {\bibinfo
  {journal} {Phys. Rev. C}\ }\textbf {\bibinfo {volume} {13}},\ \bibinfo
  {pages} {1226} (\bibinfo {year} {1976})}\BibitemShut {NoStop}%
\bibitem [{\citenamefont {Baran}\ \emph {et~al.}(2011)\citenamefont {Baran},
  \citenamefont {Sheikh}, \citenamefont {Dobaczewski}, \citenamefont
  {Nazarewicz},\ and\ \citenamefont {Staszczak}}]{Bar11}%
  \BibitemOpen
  \bibfield  {author} {\bibinfo {author} {\bibfnamefont {A.}~\bibnamefont
  {Baran}}, \bibinfo {author} {\bibfnamefont {J.~A.}\ \bibnamefont {Sheikh}},
  \bibinfo {author} {\bibfnamefont {J.}~\bibnamefont {Dobaczewski}}, \bibinfo
  {author} {\bibfnamefont {W.}~\bibnamefont {Nazarewicz}}, \ and\ \bibinfo
  {author} {\bibfnamefont {A.}~\bibnamefont {Staszczak}},\ }\href {\doibase
  10.1103/PhysRevC.84.054321} {\bibfield  {journal} {\bibinfo  {journal} {Phys.
  Rev. C}\ }\textbf {\bibinfo {volume} {84}},\ \bibinfo {pages} {054321}
  (\bibinfo {year} {2011})}\BibitemShut {NoStop}%
\bibitem [{\citenamefont {Reinhard}\ \emph {et~al.}(1984)\citenamefont
  {Reinhard}, \citenamefont {Friedrich}, \citenamefont {Goeke}, \citenamefont
  {Gr\"ummer},\ and\ \citenamefont {Gross}}]{RFGGG84}%
  \BibitemOpen
  \bibfield  {author} {\bibinfo {author} {\bibfnamefont {P.~G.}\ \bibnamefont
  {Reinhard}}, \bibinfo {author} {\bibfnamefont {J.}~\bibnamefont {Friedrich}},
  \bibinfo {author} {\bibfnamefont {K.}~\bibnamefont {Goeke}}, \bibinfo
  {author} {\bibfnamefont {F.}~\bibnamefont {Gr\"ummer}}, \ and\ \bibinfo
  {author} {\bibfnamefont {D.~H.~E.}\ \bibnamefont {Gross}},\ }\href {\doibase
  10.1103/PhysRevC.30.878} {\bibfield  {journal} {\bibinfo  {journal} {Phys.
  Rev. C}\ }\textbf {\bibinfo {volume} {30}},\ \bibinfo {pages} {878} (\bibinfo
  {year} {1984})}\BibitemShut {NoStop}%
\bibitem [{\citenamefont {Negele}(1989)}]{NJ89}%
  \BibitemOpen
  \bibfield  {author} {\bibinfo {author} {\bibfnamefont {J.~W.}\ \bibnamefont
  {Negele}},\ }\href
  {http://www.sciencedirect.com/science/article/pii/0375947489906763}
  {\bibfield  {journal} {\bibinfo  {journal} {Nuclear Physics A}\ }\textbf
  {\bibinfo {volume} {502}},\ \bibinfo {pages} {371} (\bibinfo {year}
  {1989})}\BibitemShut {NoStop}%
\end{thebibliography}%
%\bibliography{../../../information/refs/JabRef/sgzhou}

%\end{thebibliography}%

\end{document}